# Islamic Law, Western European Law and the Roots of Middle East's Long Divergence: a Comparative Empirical Investigation (800-1600)[*]

Hans-Bernd Schäfer         Rok Spruk

**Abstract**

*We examine the contribution of Islamic legal institutions to the comparative economic decline of the Middle East behind Latin Europe, which can be observed since the late Middle Ages. To this end, we explore whether the sacralization of Islamic law and its focus on the Sharia as supreme, sacred and unchangeable legal text, which reached its culmination in the 13th century, had an impact on economic development. We use the population size of 145 cities in Islamic countries and 648 European cities for the period 800-1800 as proxies for the level of economic development, and construct novel estimates of the number of law schools (i.e. madāris) and estimate their contribution to the pre-industrial economic development. Our triple-differences estimates show that a higher density of madrasas before the sacralization of Islamic law predicts a more vibrant urban economy characterized by higher urban growth. After the consolidation of the Sharia-based sacralization of law in the 13th century, greater density of law schools is associated with stagnating population size. We show that the economic decline of the Middle East can be partly explained by the absence of legal innovations or substitutes of them, which paved the way for the economic rise of Latin Europe, where ground-breaking legal reforms introduced a series of legal innovations conducive for economic growth. We find that the number of learned lawyers trained in universities with law schools is highly and positively correlated with the Western European city population. We also show with counterfactual estimates that almost all Islamic cities under consideration would have had much larger size by the year 1700 if legal innovations comparable to those in Western Europe were introduced. By making use of a series of synthetic control and difference-in-differences estimators our findings are robust against a large number of model specification checks.*

# 1 Introduction

During the first centuries of Islam's existence, economies and societies in the Middle East and North Africa were characterized by a level of prosperity unparalleled in Western Europe, by high rates of literacy and blossoming intellectual life (Pamuk and Shatzmiller 2004, Shatzmiller 2022). Nearly three centuries after its foundation, Baghdad evolved into the largest city worldwide, boasting a blossoming cosmopolitan urban life, the global zenith of culture and commerce, and the worldwide center of scientific learning (Heidemann 2006, 2010, 2014, Romanov 2017). At the same time, Western Europe in the early Medieval Period could hardly be distinguished from a social and economic backward. It was ravaged by Germanic invasions and the decline and fall of the Roman Empire. Economies shrunk to village-based self-sufficiency and city population declined. Around the year 1200, trends in both regions changed in the opposite direction and the changing trends persisted for many centuries to come, which inevitably led Western Europe to overtake Islamic countries during the 17th century. Most indicators of economic and social development suggest that both North Africa and Middle East had fallen behind, either in terms of per capita income, wages or human development. Even today, North African and Middle Eastern societies have lower per capita incomes, lower longevity, reduced rates of human capital investment, more exclusionary political and social institutions, and are more prone to a prolonged instability and civil wars than their European peers (Pamuk 2006, Kuran 2023, Keseljevic and Spruk 2023). The historical mechanisms explaining such long-haul and persistent reversal of fortune between Western Europe and the Middle East is a source of vibrant scholarly debate (Acemoglu et. al. 2002, Kuran 2004, Austin 2008).

The notion that societies in the long run cannot flourish without the institutional environment that supports economic growth and human development can seldom be disputed (North and Weingast 1989, Acemoglu et. al. 2005, Mokyr 2009, Cooter and Schäfer 2013). Legal institutions were pivotal in the medieval expansion of human capital, laying the seeds of late medieval jump-start of the Western European economies (Greif 2006, Cantoni and Yuchtman 2014, Rubin 2017, Wahl 2019). Against this backdrop, starting in Bologna and Paris in the 11th and 12th century, universities served as the centers of scholarly learning where law emerged as an independent scientific discipline (i.e. *Verwissenschaftlichung*) conducive to the rise of commerce and economic growth (Berman 1985, Schäfer and Wulf 2014, Spruk 2021).

Compared to the early medieval European economy, Islamic caliphates were known for prudent maintenance of law and order, which allowed early Islamic rules to impose low-cost economic institutions for commercial exchange that bolstered predictability, reduced transaction costs and in an empire of huge dimensions facilitated commercial interactions over vast distances. In this respect, early Islamic empire benefited greatly from a homogenous language and culture, and efficient caravan-based trade network with few barriers to trade and enforcement of contracts within the empire (Bosker et. al. 2013).

The demise of this prosperous phase of Islam is subject of much disagreement behind the critical juncture marking the beginning of decline. Diverse explanations have been proposed such as the emphasis of anti-scientific discourse by influential philosophers (Mokyr 2018) that has been later questioned by a notable number of scholars (Hassan 1996, Saliba 2007, El Rouayaheb 2006, 2015); political fragmentation and the associated territorial losses with increasing internal struggles (Bosker et. al. 2013), and the increasing political power of conservative elites precipitating a decline in scientific output (Chaney 2023). We explore whether changes of the legal framework contributed to this unfavorable development. We examine the contribution of Islamic law schools to pre-industrial economic development of the Middle East and North Africa compared to Western Europe for the period 800-1600. To this end, we employ structural break tests and identify the beginning of the 13th century as the period marking the decline and end of growth of Islamic cities. During the same time city growth in Western Europe accelerated.

**Figure 1**: Average number of residents living in cities in Western Europe, Middle East and North Africa (in thousands) from 800-1800

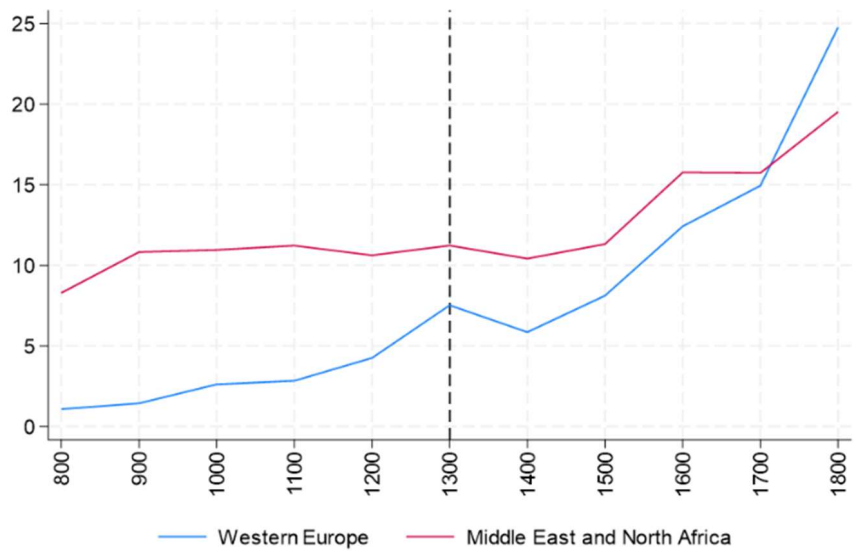

Figure 1 shows the average size of city population from 800-1800 in countries of Latin Europe and in Islamic countries of North Africa, the Arab Peninsula, the Middle East and Anatolia. For these regions and periods estimates of city population exist (Bosker et. al. 2013). With regard to Islamic countries this implies that our study does not include Iran as well as Central and South Asian Islamic countries, for which such estimates do so far not exist. We use these data as proxies for economic development throughout this study. They inform about the size of an agricultural surplus, with which a non-rural population can be fed, and therefore indirectly about agricultural labor productivity, about the level of specialization of production and trade, but also about conspicuous consumption of a ruling class. Capital cities or residential cities of a ruler might serve less well for the purpose. The graph (i.e. Fig. 1) shows that in 800 and later the Islamic region had a much higher level of economic development in comparison with Latin Europe. The application of Zivot and Andrews (2002) structural break test to the city population series for the period 800-1800 shows that a structural break exists for both groups of countries in the year 1200, but in different directions, city growth slows down significantly in Islamic countries and increases significantly in Latin Europe. We relate this break to the legal order and their change in both regions.

We present a novel dataset that tracks the evolution of Islamic law schools (i.e. madrasas) across the entire Western Islamic empire from the period 800 onwards. By employing a series of triple-differences and synthetic control-based empirical strategies, we exploit the variation in the foundation of law schools before and after the salient structural break in 13th century as a source of variation in pre-industrial economic development by levering city-level urban development trajectories in Middle East and North Africa against Western European benchmark. A series of heterogeneity-robust, city-level difference-in-differences and synthetic control estimates highlights the foundation of law schools after the legal changes of the 13th century as a source of delayed adoption of growth enhancing institutions and subsequent underdevelopment of the Middle East and North Africa compared to Western Europe. Whilst universities with law faculties in Latin Europe acted as a catalyst for the growth of new legal forms in all areas of the law we show that no such relationship between madrasas and city growth exists for Middle Eastern and North African cities over a long period prior to the start of the industrial revolution. The estimated negative impact of law schools to the city-level pre-industrial urban development survives a large battery of specification checks and placebo analyses.

Our contribution is several-fold. First, we present and examine the hypothesis that the increasing focus of Islamic law on the Sharia, which culminated in the 13th century, which different scholars label as closing the door of ijtihad, as Sharia turn (Müller 2022), as conservative originalism or as Shariatic turn contributed to the long-term pre-industrial economic development of the Middle East and North Africa. Second, by using both basic and more advanced techniques of difference-in-differences and synthetic control-related identification strategies, we estimate the missing counterfactual scenario associated with these legal developments, and find large-scale losses in terms of foregone urban development until the early 17th century. Third, by using city-level population size estimates as a rough and indirect proxy for pre-industrial development, we are able to assemble a large matched dataset that tracks the evolution of Islamic law schools and city-level population trajectories, allowing us to exploit the foundation of law schools as a source of quasi-experimental variation in urban development trajectories between Middle East and North Africa and Western Europe. And fourth, we propose and refine a disaggregated data-driven identification of the structural break date marking the end of the upward economic development, which some scholars name the Islamic Golden Age in the potential presence of unit root without any prior parameters and with the exact distribution of breakpoint test statistics.

The rest of the paper is organized as follows. For the purpose of hypothesis building section 2 presents the legal changes between Latin Europe and the early Islamic world from the early to the late Middle Ages. Section 3 presents data and samples. Section 4 presents the identification strategy. Section 5 presents and discusses results and the associated robustness checks in depth. Section 6 concludes.

## 2  Comparative legal changes in Latin Europe and Islamic countries

### 2.1  Legal developments in Latin Europe and Islamic societies during the Middle Ages

#### 2.1.1  A quiet cultural revolution in Latin Europe (1100-1300)

In this section we survey findings of legal historians on the development of law and legal norms in Western (Latin) Europe and in Islamic countries in the middle ages, which interests us for hypothesis building.

Some Roman law had survived in the kingdoms evolving on the territory of the West Roman empire in Europe and in the East Roman empire, but in a form which was hardly conducive for economic growth. Roman law, however, did not disappear completely. It remained alive in Italy, around the Gulf of Lyon and parts of Iberia. These states continued to use legal practices, which had evolved during the late antiquity, when the principate, introduced by emperor Augustus, had changed into a "dominate". The principate was still based on republican roots with limited powers of the emperor and allowed a flourishing urban economy. The dominate of late antiquity destroyed this backbone of the Empire's wealth by a ruthless system of taxes, services, regulations and requisitions for the main purpose of maintaining the bureaucracy and the army (Wieacker 1981, p. 273)

After the migration period (Germanic invasions), in almost all of Europe the predominant law became customary law of Germanic tribes. Even though during the late antiquity and early

Middle Ages the Roman legal heritage had still some practical importance, these laws, (for instance Salic law of the Frankish kingdom) concentrated rights and duties at the community level. They demanded clan loyalty, solidarity, and fellowship and had no elaborated concept of individual will or responsibility (Wieacker 1995, p. 17).

This changed fundamentally during the 12th and 13th century by legal innovations, which Gordley (2013, p. 28) called a "big bang" and Berman (1983, p. 99) a "legal revolution". The focus of these changes was the concentration of legal norms on the individual, mainly based on theological and philosophical theories. Several catholic councils and reformist popes from Gregory VII (1073-1085) to Innocence IV. (1243–1254) promoted this, influenced by writings of Anselm of Canterbury (1033-1109) to Thomas of Aquino (1225-1275). They taught that a meaningful life depends on individual choices more than on following custom or being part of a family or clan (Repgen 2014). Recognition of the individual will and private autonomy, above all the freedom of contract became a driving force for far reaching legal reforms, which were realized within a few generations. All major changes could be consolidated for the long run (Behrman 1983, p. 99).

Reform popes promoted a science based and systematized collection of papal decrees as a reformed canon law adjudicated in ecclesiastical courts. They were under the jurisdiction of the pope for all matters of marriage law and law of succession. In addition, the jurisdiction of canon courts also included the adjudication of other offenses qualified as 'sinful actions', such as usury, perjury, breach of oath or assault on clergy and churches (Thier 2009).

The resurrection of Roman (Justinian) civil law in Latin Europe since the end of the 11th century affected canon and secular courts. Roman law concentrated rights and obligations at the level of the individual and infiltrated or replaced customary law of Germanic tribes, which was family, clan, and community oriented (clan liability, clan revenge) and stressed community values as opposed to individual interests and private autonomy, which dominated Roman law and even more so canon law. This was a gradual and long-lasting process (Wieacker 1995, Whitman 2003).

Universities with autonomy from rulers and from the church became centers of higher education, especially of legal education. A non-clerical intellectual class gradually crowded out traditional clerical elites (Schumpeter 1965, p. 122; Lück 2017; Mousourakis 2015, p. 237). Judges increasingly were "learned lawyers" trained in universities. From the 13th century onwards also canon courts were filled with university graduates. The discipline of "learned law" became international within Latin Europe, with the same curricula, including Roman and canon law in all Western European universities, including those in England.

Scholastics developed a science of law for the content of private (civil) law, mercantile law, urban law, manorial law, feudal law. The science consisted of the establishment of a systematic and consistent "legal order" with rules of interpretation, hierarchization of legal norms, solving conflicts of laws, proposing overarching legal principles, and using analogy and teleological reasoning. This gave law a new flexibility and innovativeness. In the words of Zimmermann (2016, p. 332):

> *"A characteristic feature of this tradition is the pursuit of rationality, a scientific character of law, intellectual coherence and system, while ensuring organic capacity for development"*

We conjecture that these legal changes, which had their origin in the teachings of law scholars and university professors during the late Middle Ages, had a positive impact on the long run economic development in Western Europe.

### 2.1.2   Islamic law until the 13th century

During his lifetime (until 632), the head of the Islamic community modified existing tribal law by referring to the current revelation. At the time the Koran had not a completed written form and was compiled from different sources three decades after the prophet's death. Mohamed's successors as leaders of the Islamic community, the caliphs, who ruled a large Islamic empire, which stretched from Spain to Central Asia, ultimately decided on questions of Islam by drawing on early Islamic references. But they did not regard the Koran a formal source of law (Müller 2022, p. 67). In the period of the Islamic empire ruled by califs up to the middle of the 9th century law was to a large part Caliph law by decrees. Legitimacy came not from the Koran but from the status of the ruler as a representative of god (Müller 2022). During this period of Caliphate, independent jurists had limited but visible influence. Legal discussions among independent jurists and legal scholars, which influenced court decisions, seldom included references to the Koran (Müller 2022). Influences came from classical Greek philosophy, from Roman, Jewish, and pre-Islamic Iranian law and co-existed. They formed legal opinions and influenced court decisions. This practice of a pluralistic but not systematized law led to ambivalence and legal uncertainty. (Müller 2022, Oberauer 2022, Oberauer 2021, Coulson 1964, 2011).

With the end of empires, which span over the territory of the Islamic culture the importance and legitimacy of a law set by the caliphs diminished in favor of law shaped by the opinions independent jurists. The territorial fragmentation of Islamic countries did not lead to fragmentation of law but to its sacralization. Jurists systematized, developed and solidified Islamic law for the entire Islamic culture. From the 9th century onwards, the independent jurists claimed to replace the Califs as heirs of the prophet and interpreters of the religious texts. Jurists presented a systematized Sharia related body of law that transcended the legitimate power of rulers and was applicable in all Islamic countries. Schools of law (madrasas) with an interregional influence gained control on the jurisdiction of courts. The caliphs lost control on the interpretation of Islam (Müller 2022, p. 79).

The Sharia (divine instruction) consists of law related texts of the Koran and the Sunna. The latter is a collection of sayings of the prophet on morality, law and precedence, handed down by contemporaries and close followers. Legal scholars collected and compiled the Sunna texts only in the 9th century. It is debated whether they are reliable sources or reflect later views. A sub collection of such texts was canonized and received the status of a sacred tradition.

Since the 9th century more and more schools of law (madrasas) emerged, in which Sharia law was taught. Their teachers produced fundamental works and manuals and sharpened the foundations of law. Islamic law underwent significant development. It evolved into a comprehensive legal system with various schools of thought, each offering its interpretation of Sharia. Prominent among these schools were the Hanafi, Maliki, Shafi'i, and Hanbali schools, each with its approach to interpreting and applying Islamic law. This diversity allowed for adaptability and relevance across a wide geographic expanse. A multiplicity of opinions, ranging from more liberal to more conservative or originalist ones, competed. Even though the Sharia became a focal point of legal reasoning until the 13th century the law of jurists and the Sharia law still remained separated. The Kadi office became emancipated from the ruler and mutated to an autonomous subsystem. But Sharia law still remained separated from other fields of law that is the law of jurists.

In the 13th century another fundamental change occurred, which Müller (2022, p. 96) names Shariatic turn (i.e. *schariatische Wende*), also called Sunnitic revival. It denotes the expansion of the Shariah to the supreme law, which governs all of law, melted Sharia law and the law of jurists, and included a sacralization of all law under the new term "purified Sharia". The division of legal opinions which were directly related to the Sharia and other legal opinions disappeared and all ruling legal opinions became shariatic. Even the sequence of acts necessary for the making and performance of a sales contract could now be religiously interpreted as a ritual for avoiding sinful behavior and was then not subject to individual choices. (Oberauer 2021, p. 75). From then on one can speak of "Sharia law" penetrating all of law. From the 13th to the 19th century, the term "validity" in legal documents gets sweepingly replaced by "Shariatic validity" (Müller 2022), marking a change of paradigm of the basics of now a fully sacralized law. The four competing legal schools continued to exist and to arrive at different results, but the formerly important differentiation between Shariatic indicators and rational indicators for finding the law disappeared. This Shariatic turn happened within a short period of the 13th century in a time of political turmoil, extreme destructions and conquests of Mongolian invaders, after which, the city of Baghdad fell in 1258. The coincidence of these events and the turning to an all-pervading piety and submission to the law of God was perhaps more than mere coincidence. The rules of the purified Sharia law were more difficult to criticize, replace, modify or supplement than in the centuries before, which led to criticism of a fossilization of law. Older writings use the term "closing the gate of ijtihad" followed by centuries of legal stagnation. Ijtihad stands for a rational, teleological, and public welfare-oriented interpretation of the law. Coulson (1964, p. 2) writes: "*the Shari'a, having once achieved perfection of expression; was in principle static and immutable.* "*It presented the eternal valid ideal to which society must aspire*" ... and "*the notion of historical process was wholly alien to classic Islamic jurisprudence*" (Coulson 1964, p. 3).

Many present-day scholars of Islamic legal history reject this verdict. Sharia law during the 13th to 19th centuries was open to change. It took real-world influences and custom into account when evaluating applicable legal rules, which also included socially motivated assessments (Müller 2022, p. 417 and p. 453). Müller stresses that with the canonization of the Sharia basic books on law which expressed a consensus were cited for centuries to come. But inevitably, over such long periods the semantic of words and consequently their interpretation change. Therefore, the very idea of an immutable law might be inconsistent. Still, it seems that the flexibility of law was reduced after its sacralization in the 13th century. (Oberauer 2021, p. 1) argues that "*that canonization made a significant contribution to stabilizing legal doctrine and thus to conferring an element of rigidity*". (Oberauer, 2021,I). Müller (2022, p. 6) criticizes the widespread contemporary rejection of a fossilization of Islamic law in the 13th century, without offering a convincing and comprehensive alternative and argues that the modification of Shariatic casuistic rules was highly complex and needed exceptional jurists (Müller 2022, p. 336). Müller shows that the law after the Shariatic turn remained flexible, responded to social and economic demands and accepted custom. But he compares the legal development of the Sharia law from the 13th to the 19th century with the dynamics within the Mandelbrot set, in which a point in a complex number space changes coordinates by every step in a series of iterations but remains always in the vicinity of the starting point. It cannot break out and remains in the confines of a bounded set, whose shapes are characterized by self-similarity.

In the next section we show that in comparison with the legal changes, which occurred in Latin Europe at the time of the Shariatic turn, Islamic law did not develop the legal institutions necessary for developing a comparable speed of economic development afterwards.

*2.2    Legal changes in Latin Europe in comparison to Islamic countries (1100-1300)*

### 2.2.1 Enforceability of all contracts

In classical Roman law only a numerus clausus of contracts was legally enforceable. In the 13th century canon courts held that all promises must be kept, that breaching a promise is a sin. This abolishment of the numerus clausus in canon courts spread out to secular courts. All contracts became enforceable (*pacta sunt servanda*). This was a fundamental change of the civil legal system. It allowed individuals the invention and development of formerly unknown because unenforceable contracts. The insurance contract was one of those with a huge economic importance especially in the transport industry. (Gordley 2013, p. 64; Thier 2009). In Islamic law of the time a numerus clausus of contracts existed too. Courts showed some flexibility with regard to the scope of existing and the introduction of new contracts (Oberauer 2022). But a reform, which made all fair contracts enforceable, did not evolve. A lease contract for agricultural plants is observable since the 12th century. Stipulations were also possible. Such changes were marginal in comparison with the developments in Latin Europe. The Sharia rejects gambling, and a century long interpretation of the Sharia regarded insurance contracts as gambling.

### 2.2.2 The use of legal person for organizing business

The legal person existed in the period of Emperor Justinian for charities but not for profit-oriented companies. The established and predominant form of a business was the partnership, in which each partner was the owner or co-owner of the company's assets. This changed in the medieval reform period in Latin Europe. The legal person with the right to buy, own and sell things, receive gifts, have standing in court by representatives, was adapted to business purposes with new forms of contractual business charters (Berman 1983, p. 220). The corporation became systematized with two features of far reaching importance. It has the legal capacity to act, and rights and duties of the corporation became distinct from those of its shareholders. This implied two important economic consequences (Hansmann and Kraakman 2000).

First, it provides an asset shield of the corporation against creditors of shareholders, who fail to service their private debts. In a partnership such unreliable partners expose the firm to existential risk because the creditor of a partner, who does not pay his personal debts, can seize the assets of the company, which exposes the firm to existential risk. The reason is that assets in a company are organized as a net unlike cash money. If the creditor can seize the transport fleet of a company all other assets might be worth nothing. Organized as legal person the company itself owns the assets, and the creditor of a defaulting shareholder can only seize the shares but not the company's assets. In a partnership each partner must therefore closely monitor the reliability and financial capability of every other partner. Consequently, the number of partners remains small. In a company, which is a legal person, this is not necessary. The personal financial situation of a shareholder is of little or no interest for other shareholders as soon as the shareholder paid for his company share. Many shareholders, big companies, and stock markets become possible (Kuran 2003, 2016).

Second, the corporation as legal person exists for an indefinite period. Unlike in a partnership the death of a shareholder does not expose the business to existential risk. The heirs inherit the shares and not their part of the assets of the company as in a partnership. This removes the risk that a company disappears with the death of one of its partners. Both effects taken together allowed the existence and stability of large companies with many shareholders, which could exploit economies to scale and exist for long time spans in Western Europe. In Islamic countries, a comparable legal innovation did not occur. An institution with the characteristics of a legal person existed for foundations (*waqf*). It allowed the creation of long-lasting organizations, which could finance public infrastructure, mosques and centers of higher education (madrasas) (Müller, 2022, p. 416). It was never contractually transformed into a corporate charter. The partnership remained the main business model, which seldom could grow to a size bigger than a family business (Kuran 2013). The legal person was introduced in the Ottoman empire only in the 19th century as a public company for financing the railway system.

### 2.2.3 *Systematic rejection of the ban on interest for credits versus circumventions*

Scholastics reinterpreted the canonical ban on interest and made it an almost empty shell except for consumer credits (Tan 2002). They argued that a loan, which led to "*damnum emergens*" (loss) or a "*lucrum cessans*" (lost profit) of the creditor interest payment was regarded as compensation for this loss and was then not a sin of usury. Both concepts cover a wide range of case law. They provide the impression of exceptions. However, Schumpeter writes that the medieval scholars came close to a general theory of interest, which became fully developed only in the 19th century and nothing of relevance was written in between. This supported the emergence of a finance industry, without which rapid growth of companies is hampered. Interest remained usury mainly for consumer credits (Schumpeter 1965, p. 177; Zimmermann 1990, p. 171). They became a business for Jews, which contributed to their discrimination. This judge made rejection of the ban on interest was not a circumvention, but an outright teleological rejection on the ground that interest is not usury if it is factually a compensation for a loss or for a lost profit caused by lending. Another, comparatively crude method of getting around the ban on interest was the *contractum trinius*, which sometimes led to the legally accepted factual interest on a credit by a combination of three contracts (Calder 2016). It is a contract, which makes the lender a shareholder of a company. This contract is combined with an insurance contract, which insures the new shareholder against losses of the firm and fixes the profit share to be paid per period and a third contract, which excludes a gain higher than the agreed amount of profits. Compared with the teleological interpretation and rejection of the ban this is a crude and tricky circumvention. It is noteworthy to emphasize that some courts accepted this particular construction.

The Sharia bans interest taking with sharp words of the Sunna (Oberauer 2014). An outright teleological rejection of the *riba* ban did not gain recognition. A substitute institution for a credit against interest existed, the *mudaraba*, a silent partnership, in which the partner was paid a fixed profit as a percentage of the investment level, insofar comparable with a credit. However, in case the investment got lost the partner had no claim. The *mudaraba* implied capital at risk. It was therefore not an effective substitute for a credit (Oberauer 2014). As a tricky circumvention it could also not provide the legal protection of an outright teleological rejection of the ban. Often the ban on interest was neither strictly enforced nor was the practice among merchants always in line with the legal situation. However, the *riba* ban remained an impediment to the rise of a finance industry, which depends on reliable legal protection of credit contracts, without which the growth of companies depends on self-financing. Credit contracts with a de facto interest existed in Islamic countries, but the frequency differed across time and space and it seems that a large finance industry for the benefit of companies did not develop (Oberauer 2014, p. 122).

### 2.2.4 Good faith purchase

In sales law, the possibility of acquiring mobile property from an unauthorized person in good faith was introduced in Western Europe in the late Middle Ages. This was unknown in classical Roman law. This innovation protected transactions at the risk of the rightful owner but reduced overall transaction costs and consequently increased the joint surplus from contracts. It supported the rise of extended and more anonymous open markets. (*Mobilia non habent sequelam*) (Berman 1983, p. 236). This rule did not evolve in Islamic law as the validity of the sale remained contingent on the consent of the rightful owner. This requirement removes the economic advantages of the good faith purchaser rule.

### 2.2.5 New forms in mercantile Law

The focus of the legal reforms on private autonomy allowed the development of internationally observed mercantile law, which included prototypes of limited liability, and marine insurance contracts (Berman 1983, p. 233). In Islamic countries insurance contracts did not evolve as the Sharia regarded them as gambling with fate. The obvious teleological argument that an insurance contract reduces risk and removes even existential risks was obviously not successful against the prevailing interpretation of the Sharia. In mercantile law new forms, which reduced transaction cost were introduced in the 12th century in Italy, including abstract bonds, bills of exchange and promissory notes (Wieacker 1967, p. 241, Mousourakis 2015, p. 349). Such financial papers like the bill of exchange were used in Islamic countries already in the 7th century. However, in the late Middle Ages, when they were introduced in Latin Europe the holder of the bill of exchange could transfer the bill before maturity with a discount on the nominal value and this transaction was protected by law. In Islamic countries, a discount on the nominal value was a flat violation of the *riba* ban (ban on interest), which implies the view that a sum of money must have the same value at different points of time, and that the passage of time alone cannot change its value. This implies that a discount on the nominal value of a commercial paper is sinful.

### 2.2.6  Possibility to testate

Canon law introduced the sanctity of the testator's will into the law of succession and improved the social and economic position of women (Behrman 1983, p. 233). The possibility to testate at all was recognized. This implied the possibility of an arbitrary succession. The introduction of an arbitrary last will falls in the general context of the recognition of individual autonomy. This innovation replaced the Germanic inheritance law, which was completely customary, without last will (Zimmermann 2016, p. 48). Canon law of succession also improved the position and economic capabilities of women. The surviving wife inherited a share. This share did not exist in customary law and was more secure and higher than under Roman (Justinian) law. Women played a more important role in business and politics, if the ruler of a county or kingdom was determined by the law of succession. Based on the testator's will, they could sometimes inherit a business, a county or kingdom. (On the contrary, in European republics with elected rulers as in the free cities women played no role in politics until some decades ago). In Islamic countries the last will was recognized by law, within the limits of Sharia and with fixed shares for family members including the testator's wife. A woman could not inherit a polity like a county or kingdom. Independent from the fixed shares of relatives the testator could only transfer a maximum of one third of his assets by his arbitrary last will. The rules of succession made it difficult for parents, who had no direct male offspring to transfer their assets to a close female relative. A large part went to more distant male relatives or even to the state (Müller 2022, p. 382). It seems that the Shariatic turn did neither improve nor worsen the status and capabilities of women over time. In comparison, in Western European legal developments implied an improvement of the status of women via changes of the law of succession. In Latin Europe canon law made marriage exclusively dependent on a mutual promise. All other factual influences became legally irrelevant. In Islamic law the woman had to agree to the marriage contract, except when she was married by her father or grandfather (Müller 2022, p. 366).

### 2.3  Conjecture

Islamic law gained legal certainty but lost flexibility for a long time starting with the 13th century when the sacralization of Islamic law by jurists reached its culmination in the Shariatic turn (Müller 2022, p. 41). Also, the fundamental and growth inducing legal innovations, which occurred in late-medieval Latin Europe at the same time, had no counterpart in Islamic countries. These are, (I) the replacement of the Roman numerus clausus of contracts by a new enforceability of all fair contracts, (II) the related invention and introduction of insurance contracts, (III) corporate charters, which transformed the legal person into a business organization allowing the establishment of big and long-lasting for profit companies, (IV) the systematic, theoretically informed and teleological removal of the ban on interest for commercial credits, which allowed the rise of a legally protected finance industry, (V), the good faith purchase, which supported the emergence of larger and more anonymous markets and (VI) the improved role of women in politics, society and the economy by reforms of the canon law of marriage and succession. Also, it seems that for none of these innovations there existed or developed substitute institutions with a comparable effectiveness in Islamic countries. All these differences came into existence in the late Middle Ages but were consolidated for centuries to come.

We therefore conjecture that from the 13th century onwards the law increased economic growth (proxied with the growth of city population) in Latin European counties until the year 1600 and that legal institutions had little positive and possibly a negative effect on city population in Islamic countries in comparison with Western European countries and during the same period of time.

## 3    Data and samples

### 3.1    Dependent variable

Our dependent variable denotes city size, measured as the number of inhabitants in thousands for each city. For the pre-industrial period, the size of the cities in Medieval Period provides both a reasonable and relatable indicator of economic and social prosperity, since the medieval cities primarily based their economic activities on commerce which require larger and more vibrant city markets, which can be adequately captured by population size (Lees and Hohenberg 1989, De Long and Shleifer 1993). Although more nuanced indicators of prosperity would invariably provide better and less uncertain measures of prosperity[2], such indicators are very scarce for a larger sample of cities and per se hinder a meaningful comparison both across space and time. For a sample of 621 European cities from 21 countries[3], our main source of population size estimates is Bairoch et. al. (1988) where the corresponding estimates are provided for nine different centuries during our period of investigation. Our focus comprises the full spectrum of cities to capture gradually increasing urban population (De Vries 1984, Acemoglu et. al. 2005). For the Middle East, Balkans and North Africa, our main source of city population estimates is Bosker et. al. (2013)[4] from which we collect the city population estimates for 145 cities from 16 countries[5]. For each city our time series, which runs from the year 800 to 1800 includes data for 11 Years, one for every 100 years. Figure 2 charts the development of population size over time for a full sample of European and Middle East for the period of our investigation, and reports the mean size of cities alongside 95% confidence intervals. It indicates a relatively homogenous

---

[2] De Long and Shleifer (1993, p.675) further emphasize pre-industrial city population as a broad and rough indicator of economic prosperity by noting that "*the larger pre-industrial cities of Europe where nodes of information, industry and exchange in areas where the growth of agricultural production and economic specialization had advanced far enough to support them. They could not exist without a productive countryside and a flourishing trade network.*" They also emphasize several limitations by stating: "*The correlation between economic prosperity and city size may not hold in general for the preindustrial world. The population of Tenochtital, or Beijing, or imperial Rome had more to do with the power of the networks of tribute and redistribution that underlay their respective empires than with mercantile prosperity. Such consumption-intensive 'parasite cities' to use Paul Bairoch's term, were centres of neither trade nor urban industry but instead the homes of bureaucrats and the favoured dwelling places of landlords. But the primary rural orientation of Europe's medieval ruling class meant that Europe's cities did not develop as centres of landlord consumption or of territorial administration.*"

[3] Austria, Belgium, Czech Republic, Denmark, Finland, France, Germany, Hungary, Ireland, Italy, Luxembourg, Netherlands, Norway, Poland, Portugal, Slovakia, Spain, Sweden, Switzerland, United Kingdom, parts of former Yugoslavia without Islamic rule

[4] Although additional details behind the reconstruction of the size of cities is provided in Bosker (2013) supplementary appendix, it should be noted that the size of each city in Middle East and North Africa is constructed on the basis of the estimated number of inhabitants per hectare of surface area of the medieval city. Population estimates are derived from synthesized accounts of Arabic travellers in North Africa and the Middle East on the separate estimates from the secondary literature, Baedeker travel guides, excavation maps, and assessments of the number of local mosques and public hammams to recover both physical data and population estimates to indicate the number of inhabitants. The basis of the assumption on city size used by Bosker et. al. (2013) is 150 inhabitants per hectare of surface area or the medieval city except for the "garden" cities such as Baghdad, Basra and Sana'a where 75 inhabitants per hectare are assumed. For the cities in Arab Peninsula, the population estimate is derived from map-based indication of walled surface areas combined with separate estimates of city size from the existing literature. It should also be noted that the correlation between Bosker et. al. (2013) city size estimate and previously established estimates is between +0.98 (De Vries 1986) for Western Europe, and +0.90 (Malanima 1998) for Northern Italian cities, which implies that the reconstructed city size may not be mere artefacts.

[5] Albania, Algeria, Bulgaria, Egypt, Iraq, Israel, Lebanon, Libya, Morocco, Oman, Saudi Arabia, Spain, Syria, Tunisia, Türkiye, Yemen and parts of former Yugoslavia under Islamic rule for at least one century during our investigation. We consider Spanish cities in our sample up to the period of the fall of Granada in 1492 that were under the Islamic rule during the period of our investigation

demographic development of the European cities characterized by relatively low confidence intervals against a more diverse trajectory of development in the Middle East characterized by higher variance, and larger confidence intervals. Figure 4 contrasts the disparities in population size with the proxy estimates of per capita GDP in Western Europe, southern Iraq and lower Egypt (Pamuk 2006, Pamuk and Shatzmiller 2014), and shows that a considerable stagnation of population size corresponds with a prolonged economic stagnation beginning after the 12th century. It should be noted that the respective city size captures the size of the consumable class as a rough indicator of the level of prosperity in the medieval period. Despite the relatively broad coverage of the data in space and time, city size estimates in our treatment sample are available for the Middle Eastern cities, and not for other cities where Islam was established in South and Central Asia.

**Figure 2**: Population size and urban development in Europe and the Middle East, 800-1800

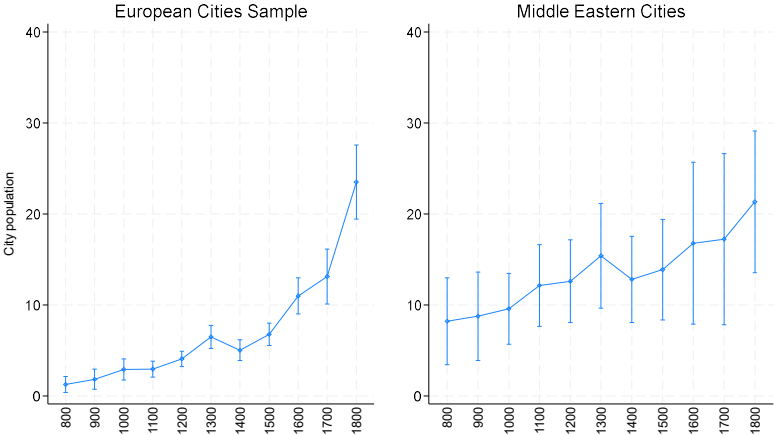

Figure 3 presents the aggregate city size trajectories for the four regions in our treatment sample (i.e. Arab Peninsula, Levantine, North Africa and Ottoman-Anatolia) along with the 95% confidence intervals. Decomposing the overall trajectory of demographic development allows us to better grasp the differences in the evolution of city size across a vast territory over a long period of time. The descriptive evidence reinforces the pattern of the relatively vibrant city growth prior to the beginning of the 13th century. The growth process appears to be somewhat more dynamic in cities on the Arab Peninsula and North Africa. Although all four groups of cities can be readily characterized by the stagnating city size beginning in 13th century, the development of cities appears to be somewhat more dynamic prior to the turn towards Sharia-based interpretation of law.

**Figure 3**: Breakdown of population size and urban development

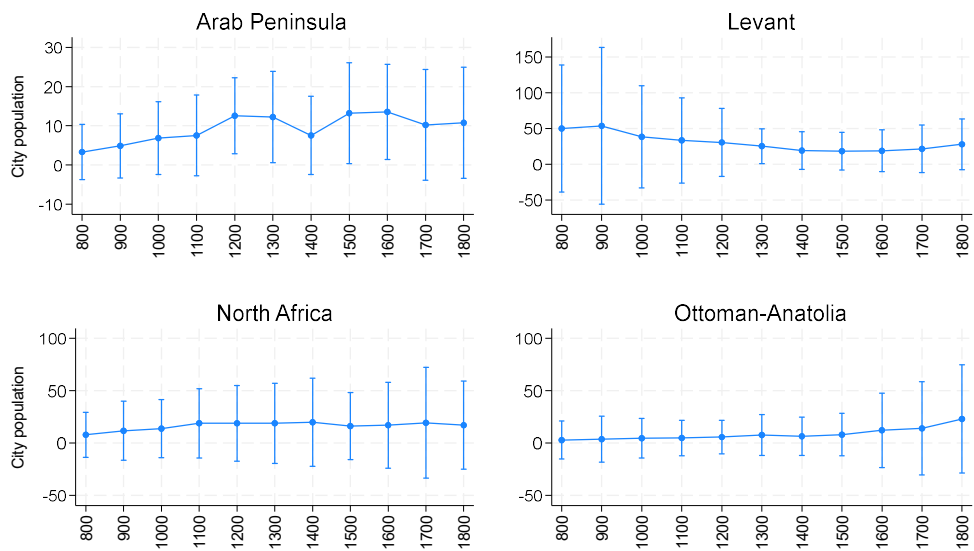

**Figure 4**: Economic growth trajectories in Middle East and Western Europe, 1-1820

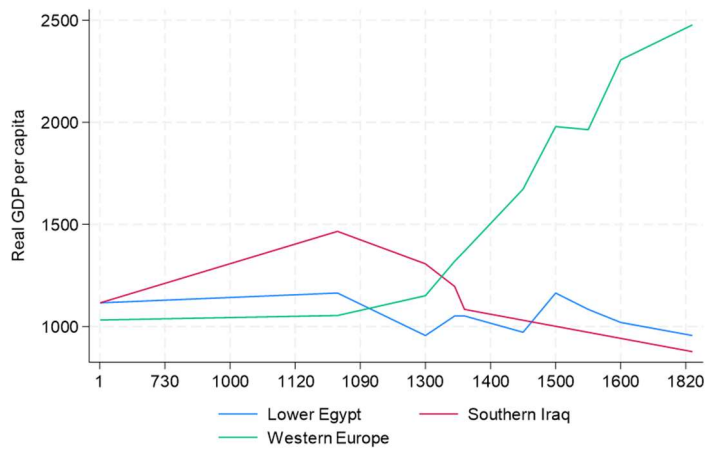

*3.2   Treatment variables*

*3.2.1   Universities and madrasas*

The university variable is a binary indicator of whether the city has had a university in a given year. We use the university variable to capture the degree to which modern learned law influence each city over time. This variable is based on the list of 148 European universities founded in the period between 1000 and 1600. For the period 1000-1500, we obtain the information on the foundation and presence of the university from Verger (2013) and Frijhoff (1996). Since not all universities had law faculties, we eliminate those without law faculties based on careful screening of the literature. But since it cannot be entirely excluded that some universities provided legal education and that cities benefitted from legal education in adjacent areas, we follow Schäfer and Wulf (2014), and construct a variable indicating the number of universities in 300 km radius, and construct an index that reflects the sum of universities within the radius. Since the universities of Bologna and Paris were particularly more important than others in legal education, we increase the value of the index by one if these universities belong to the 1,500 km radius of each city. The choice of 300 km and 1,500 km radius is based on the consideration that students of canon and civil law could study abroad at more prestigious universities in Bologna and Paris, or at a regional university closer to their hometowns. During the 12$^{th}$ and 13$^{th}$ centuries, few universities existed across Europe and this entailed a high cost of travel abroad to obtain legal education, rendering study of law available only to a small group of privileged students. By 14$^{th}$ and 15$^{th}$ centuries, more universities were founded, making it easier to obtain legal education across more regions in Europe, which gradually expanded the number of students that could feasibly obtain university education, as well as the subsequent number of qualified lawyers.

For the Islamic and Middle Eastern cities, we construct a variable indicating the number of madrasas in a given city and year. By systematically analysing a vast scholarly literature on education in Islamic societies from the initial year of our sample period, we calculate the number of madrasas for each city-year combination, and exclude those madrasas that provided no possibility of legal education. Our primary source for the number of madrasas is *Brill Encyclopedia of Islam*. For each Islamic and Middle Eastern city in our sample, we screened the existing legal, educational and architectural literature and charted the number of madrasas over the years from 800 onwards. The estimated number of madrasas is further verified by comparing our series with Bosker et. al. (2013) where a binary indicator of madrasas presence is presented. Similar to the universities in Europe, not all madrasas had faculties or colleges of law. This implies that it cannot be excluded that some cities were indirectly affected by legal education in adjacent cities since the medieval practice of frequent travel to the colleges elsewhere to obtain a legal education was particularly common. To this end, we construct a variable indicating the number of madrasas in 300 km radius of the city, and construct an index reflecting the sum of madrasas within the radius. This also reflects the network and supply of law schools providing legal education in the adjacent distance of the respective cities. Figure 5 depicts the density of madrasas across our sample of Islamic cities. It presents the overall number of madrasas as the unweighted average across the Islamic cities during our sample period. For instance, madrasas appear to have been concentrated in specific cities of importance such as Constantinopole, Damascus, Cairo and several others whilst not each city had a madrasa. Furthermore, Figure 6 reports the spread of madrasas over time across four major geographical groups of our treatment sample, and confirms an reasonably strong expansion of the madrasas network after the Shariatic turn in the early 13$^{th}$ century.

**Figure 5**: Density of madrasas in the Middle East, Balkans and North Africa, 800-1800

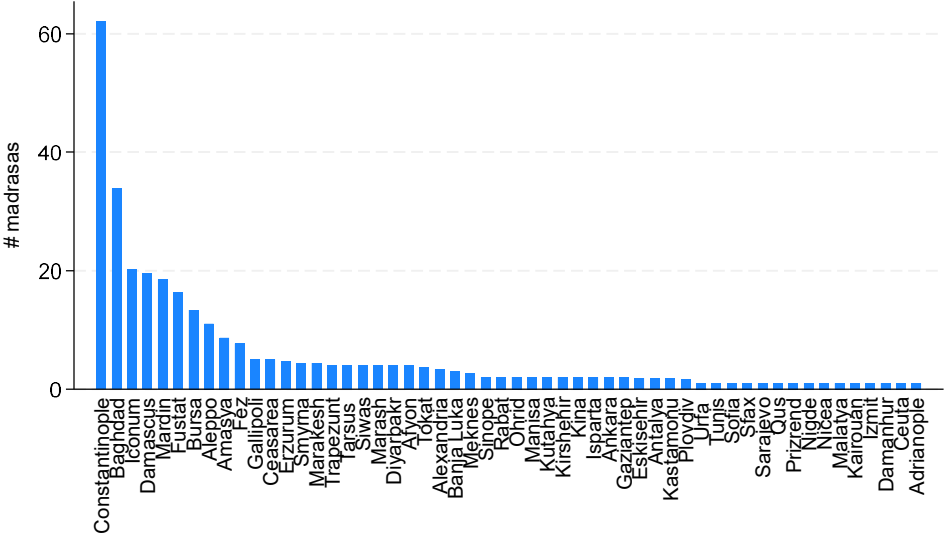

**Figure 6**: Gradual expansion of Islamic law schools across Middle East, Balkans

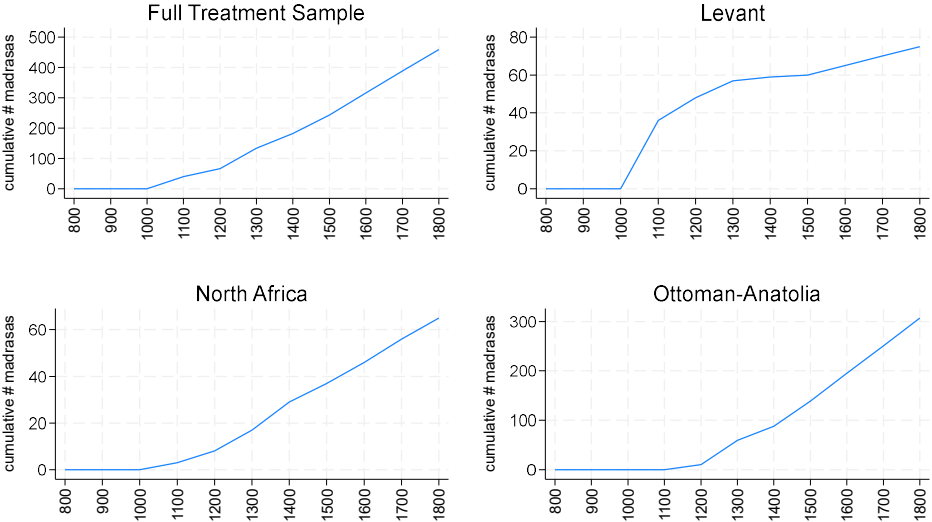

To capture the religious importance of the cities, we consider three distinctive variables. First, we construct a dummy variable indicating whether a city is a capital city (McEvedy 1977a,b) or a bishopric and archbishopric city for each century in our sample period primarily relying on the Jedin et. al. (2004) as the main source. Such cities attracted both economic activity and people based on the presence of sovereign powerholder and his or her court. In turn, cities provided services in return for taxes and rent. For the sample of Islamic cities, we capture the institutional importance of the city through a binary variable indicating the capital cities. Second, the religious importance of cities across Islamic societies is captured by the distance from Mecca. And third, the unique development of cities in the Iberian Peninsula since the conquest by Umayyad dynasty in the late 8$^{th}$ century is captured by a separate dummy variable indicating whether the cities were integrated into the Emirate of Cordoba, and later into the Emirate of Granada (Bosker et. al. 2013).

*3.3  Control variables*

*3.3.1  Institutional variables*

*3.3.1.1 Active parliament*

We capture the institutional quality of the cities in our model through two distinctive variables that confront two independent sources of variation in pre-industrial urban development. Our first variable captures the presence of active parliaments. In the seminal contribution, North and Weingast (1989) emphasize the rise commitment-based parliamentary institutions during the Glorious Revolution as one of the most salient critical junctures facilitating rapid economic takeoff of England in the 18$^{th}$ century. Nevertheless, prior literature has neglected the parliamentary activities in medieval Europe which could be partly pivotal for the pre-industrial economic growth. To fill the void in the literature, Van Zanden et. al. (2012) develop a quantitative parliamentary activity index charting the development of European parliaments between 1100 and 1800 and establish several noteworthy insights. For instance, parliamentary activity gained considerable foothold during the period of Reconquista in Spain, particularly between 12$^{th}$ and 13$^{th}$ century when the Spanish rulers cultivated closer relationship with the cities and their citizens reconquered from Muslims. They suggest that parliamentary diffusion rapidly spread across the rest of Europe and gained substantial institutional strength in Northern

Italy whilst spreading slowly to the north of Europe. In addition, the institutional divergence evolved within Europe in the early modern period when parliaments lost their significance in southern and central Europe whilst becoming important and powerful in United Kingdom, Netherlands and Sweden, possibly reflecting the institutional impetus and success of Protestant Reform in the 16th century. The presence of active parliament is captured by a binary variable indicating whether or not the city had an active parliament in place in each year under investigation.

*3.3.1.2 Political freedom*

Another related and important dimension of the institutional quality of cities is the political freedom. To this end, De Long and Shleifer (1993) constructed an eight-point scaled index of political freedom to capture the degree of political freedom.[6] We fully adopt this scale as a proxy variable for the degree of political freedom for our analysis. Nonetheless, the variation in political freedom may exhibit a strong collinearity with the parliamentary activity index, and render the estimated coefficients questionable and unreliable. To partly overcome the potential collinearity, we recode De Long and Shleifer's eight-scale classification into a binary variable that takes places 1 if the cities were independent city-republics, weak prince cities or full constitutional monarchies or republics, and 0 otherwise. Reclassifying the cities into free and unfree ones through a binary indicator partially isolates the conjuring variation in the parliamentary activity.

*3.3.1.3 Roman law*

To capture the independent contribution of the Roman law to the pre-industrial economy growth, we use Schäfer and Wulf (2014) categorial variable denoted on the range between 0 and 4, indicating the influence of substantive and procedural rules from the Justinian law as it was established and scholastically developed in Italian universities. In the simplest form, zero value indicates no influence of Roman law at all, 1 denotes little influence, 2 represents medium influence, 3 implies a high influence and 4 entails a very high influence.[7] Following the classification of influence of Roman law, for the cities within the borders of the Holy Roman Empire, our assumption is that the influence of Roman law increases over time, reaching the peak value of 4 in the year 1500 and remains stable therefrom. In 1495, the imperial chamber court for the holy Roman Empire was established, whose judges had to be trained in Roman law. It should be noted that although in the Eastern Roman Empire, Corpus Juris Civilis was compiled into a single collection during the sixth century, its spread had no effect until the beginning of the 19th century in the areas covered in our sample, specifically Greece and Balkans. Since the degree of the reception of Roman law is denoted on a categorial scale, it allows us to capture the overall impact on pre-industrial development in a more comprehensive and elaborate manner.

---

[6] The range of the index runs from military conquerors rule or full bureaucratic absolutism (0), non-bureaucratic absolutism (1), strong-prince proto-absolutism (2), feudal anarchy (3), extra-constitutional princes appointed by powerful magnates (4), independent city republics (5), weak prince rule (6), and full constitutional monarchy or republic (7).

*3.3.2    Human capital variables*

*3.3.2.1 Book production*

We also capture the contribution and influence of human capital to pre-industrial economic growth by using the variable on manuscript and printed book production (Buringh and Van Zanden 2009) as a more comprehensive and broad measure of social and intellectual capabilities such as production of ideas, level of literacy, education as well as the ability to consume more luxury goods. More specifically, the variable contains the number of individual manuscripts and printed books for the years from 1200 to 1400. For the year 1500, the variable contains manuscript count as well as the number of new titles and editions of printed books, which are multiplied by mean size of print runs. Moreover, for the year 1600, the variable contains an estimate of the total number of books and manuscripts. Although the book production does not fully reflect both intrinsic and extrinsic intellectual capabilities of the typical medieval economy, it is one of the few more reliable proxies for the literacy rates and accumulation of ideas that can be traced back to the medieval period.

*3.3.3    Black death*

---

[7] More specific details behind the geographical variation in the scope of influence of Roman law during the early and late medieval period can be found in Schäfer and Wulf (2014, p. 282, Table 2).

The Black Death was perhaps the most devastating bubonic plague in European history that haunted Europe in the period between 1347 and 1351. In spite of some uncertainty, existing scholarly estimate suggests that the overall mortality rate from the plague was between 50 percent and 60 percent, respectively (Helleiner 1967, Daileader 2007, Pamuk 2007, Kelly 2013). These estimates also invariably indicate the most devastating severity of the plague in Italy and the rest of the Mediterranean area whereas the plague had been somewhat less devastating in England, Germany and the Netherlands, possibly owing to the interplay between low population density and geographic location away from major trade and supply routes where the plague spread at its fastest pace. On the contrary, Daileder (2007) suggests that Poland and Bohemia were not much affected by the plague whilst Benedictow (2004) puts forth that Iceland and Finland are the only parts of Europe unaffected by the plague. We incorporate the confounding influence of Black Death into our model and analysis by relying on five-scale index from Schäfer and Wulf (2014) that takes five distinctive values that imply (0) no effect at all (cities in Finland), very little effect (1)[8], little effect (2)[9], moderate effect (3)[10], high effect (4)[11] and very high effect (5)[12], where it follows that higher values of the index imply a greater severity of the plague.[13] Based on the historical estimates of the extent of plague (Dols 1979), we assign the "little effect" value of the Black Death index for cities in Egypt and Syria on the same 0-5 severity scale.

### 3.3.4  *Trade and economic geography variables*

Another important driver of pre-industrial economic growth concerns the ability to undertake long-distance trade and intrinsic geographic characteristics that do not necessarily depend on the stock of existing institutions. We capture the trade and geography-related variables that simultaneously affect urban development through three distinctive variables. First, the ability to undertake near and long-distance trade depends on the potential of city's location for interacting with other cites. Such potential can be seen either as an accessibility of the location to key

---

[8] Poland and Bohemia
[9] Scandinavia, The Netherlands, Switzerland, Scotland, Ireland
[10] England, Austria, Hungary, Portugal, Germany
[11] Belgium, France, Spain
[12] Italy
[13] Furthermore, the values of the Black Death index of 1, 2 and 3 in Schäfer and Wulf (2014) correspond to the values of 1 and 2 in Pamuk (2007)

markets and trade in all other locations, or accessibility of the city to the inhabitants of other cities. That said, the potential to develop trade and foster urbanization per se depends on the physical distance of each city to the other cities. To this end, De Vries (1984) proposes a simple index of urban potential, further refined by Bosker et. al. (2013):

$$\Omega_{i,j,t} = \sum_{j \neq i}^{n} \left( \frac{M_{j,t}}{\delta_{j,t}} \right) \times \mathrm{I}_{i \cap j = \{x: x \in i \,||\, x \in j\}, t} \cdot [1]$$

where $\Omega$ is distance-weighed sum of the size of all other cities denoted by $j = 1,2 \ldots J$ for i-th city at time $t = 1,2,..T$ computed using standard latitude and longitude coordinates, $M$ is j-th city population size, $\delta$ is the respective distance from i-th city to the full set of $j = 1,2 \ldots J$ cities, and $\mathrm{I} \cdot [1]$ is the binary variable indicating any potential source of network interaction between i-th city and its j-th counterpart across $x$ element that could be either religious, institutional or any other meaningful criteria. It should be noted that urban potential reflects the ability of cities to trade and interact, and also implies that a change in population variable of each designated j-th city has a direct impact on the potential to develop trade and exchange in i-th city.

Figure 7 compares the evolution of foreign urban potential of Aleppo and Lyon for the full period between 800 and 1800. Without further empirical scrutiny, the descriptive evidence indicates a marked divergence in the foreign urban development potential. Prior to early 1200s, Aleppo is characterized by a noticeably higher foreign urban potential whereas Lyon's urban development potential tends to improve over time. In turn, this suggests a marked disparity in the degree of urban development prior to the turn towards Sharia. Afterwards, a relatively clear divergence is perceptible alongside a comparative stagnation of Aleppo's urban development contrasted against a steep rise observed in Lyon, and also elsewhere in Latin Europe.

**Figure 7**: Comparative foreign urban development trajectories of Aleppo and Lyon, 800-1800

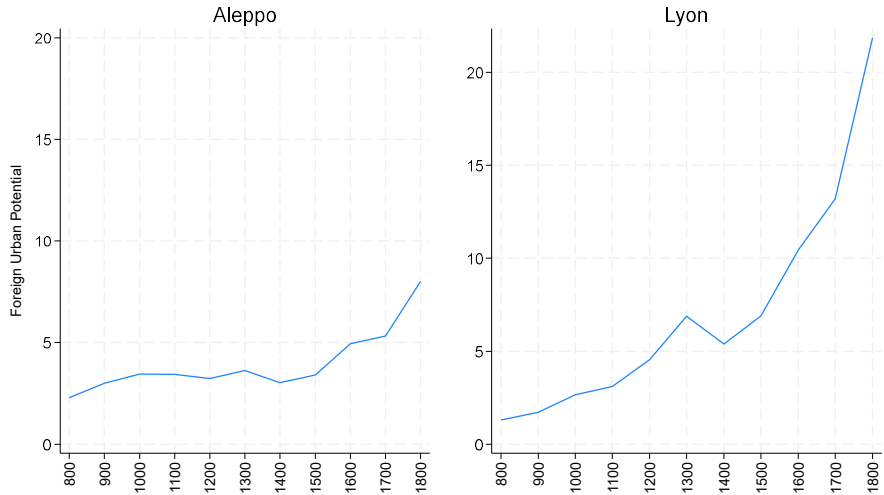

Our second variable is a simple binary indicator of access to the caravan routes which is particularly noteworthy for the Islamic cities. In the absence of easily navigable rivers, trade and exchange could only be facilitated through efficient caravan routes connected by camels. The geographic location of the Islamic empire enabled a vast pan-Islamic free trade zone further enabled by the conjuring access to three different continents. The information on the caravan routes also partly captures the potential of cities to develop agricultural potential since the location of caravan routes had to rely on the water supplies to facilitate long-distance trade. The endogeneity concerns behind the location of caravan routes may be mitigated since these locations did not shift in response to more favorable trade opportunities or institutional constraints, and remained more or less the same over time. Our main source of information on the caravan network comes from Rostovtzeff (1971), Hourani (2002) and Bosker et. al. (2013). Our third variable capture geographic characteristics that invariably matter for pre-industrial develop is a simple binary variable indicating access to the sea. This particular variable provides some information on the ability to undertake navigable water transport to embark on long-distance trade which proved to be especially pivotal in developing the Atlantic trade by late 15th and early 16th century. During this period, the access to the sea was fundamental in both colonization and institutional changes (Acemoglu et. al. 2005). Table 1 provides some descriptive statistics for both European and Middle Eastern city-level samples. Although the battery of control variables is far from exhaustive, it should be noted that the inclusion of city-specific and time-specific unobserved effects together with the interactive unobserved effects somewhat mitigates the extent of omitted variable biases that could otherwise taint the validity of our estimates.

Table 1: Descriptive statistics

|  | Obs | Mean | Std | Min | Max |
|---|---|---|---|---|---|
| **Panel A: European City-Level Sample** | | | | | |
| City population (in 000) | 6,831 | 7.631 | 24.294 | 0 (Danzig) | 948 (London) |
| Universities | 6,831 | 0.057 | 0.233 | 0 (Ocana) | 1 (Barcelona) |
| # universities in 300 km radius | 6,831 | 4.764 | 6.459 | 0 (Catanzaro) | 31 (Livorno) |
| **Panel B: Middle Eastern City-Level Sample** | | | | | |
| City population (in 000) | 1,789 | 13.207 | 40.036 | 0 (Wasit) | 700 (Constantinople) |
| # madrasas | 1,789 | 1.011 | 5.422 | 0 (Sousse) | 137 (Constantinople) |
| # madrasas in 300 km radius | 1,789 | 19.663 | 49.019 | 0 (Al-Mahalla Al-Kubra) | 400 (Beirut) |
| Emirate of Granada | 1,789 | 0.01 | 0.09 | 0 (Labda) | 1 (Velez-Malaga) |
| Distance from Mecca | 1,789 | 2944.38 | 1657.18 | 73.09 (Jeddah) | 5426 (Marakesh) |
| **Panel C: Control variables** | | | | | |
| Active parliament | 8,723 | 0.275 | 0.446 | 0 (Augsburg) | 1 (Halifax) |
| Book production | 8,723 | 23.915 | 63.441 | 0 (Schwaz) | 502.47 (Amsterdam) |
| Capital city | 8,723 | 0.038 | 0.191 | 0 (Hamburg) | 1 (Baghdad) |
| Black death | 8,723 | 0.243 | 0.967 | 0 (Danzig) | 5 (Rovigno) |
| Foreign urban potential | 8,723 | 6.623 | 6.235 | 0.664 (Galway) | 130.46 (Deptford) |
| Political freedom | 8,723 | 0.230 | 0.421 | 0 (Santarem) | 1 (Amsterdam) |
| Roman law | 8,723 | 1.108 | 1.500 | 0 (Greenwich) | 4 (Novara) |
| Caravan hub | 8,723 | 0.034 | 0.181 | 0 (Damietta) | 1 (Aleppo) |
| Sea access | 8,723 | 0.224 | 0.417 | 0 (Hannover) | 1 (La Rochelle) |

*3.3.5  Islamic law schools and treatment selection*

The next question pertains to the plausibility of the treatment selection. One of the possible avenues to assess whether or not the Sharia-based canonization (Shariatic turn) and consolidation of Islamic law may be a valid treatment-related variable is to perform the structural break test in the city population trajectories in the full sample period (i.e. 800-1800) and determine the perceptible time-specific break points. To this end, we estimate the structural break points by relying on Zivot and Andrews (2002) structural break test allowing for the presence of non-stationary growth dynamics. By estimating the full distribution of breakpoint test statistics, we are able to determine the most potent and powerful breakpoints in the time-series to inquire whether a break is perceptible in the underlying time series. Table 2 reports the structural break test statistics and the estimated dates of breaks. For Western Europe, the most significant estimated break date is in the period between 1100 and 1200, and appears to be very large (i.e. p-value = 0.000). The break date almost entirely conforms with the beginning of the cultural revolution in the 12$^{th}$ century (Berman 1985). On the contrary, the most significant structural break in the cities of Arab Peninsula appears to have taken place in the years approximately around 1300. For the Levantine cities, the estimated break date is in the time range between 1200 and 1300 whilst for North Africa, the estimated break date with most significant test statistics is precisely around the years 1200. It is perhaps worth emphasizing that the estimated time-specific break points for the cities in former Emirate of Granada differ considerably from their peer break points in other areas of the former Islamic Empire. For instance, in the cities in Iberian Peninsula, the most potent break date that takes place in years in the range between 1400 and 1500, respectively. This conforms with the gradual reconquest of the Iberian Peninsula by the Christian kingdoms and the subsequent turmoil and consolidation associated with the Castilian rule. Our evidence from structural break tests thus provides some common support for the notion that important and discernible time-specific breaks in the trajectories of city population have coincided with the cultural revolution in Western Europe and with the consolidation of the Shariatic turn during the 13$^{th}$ century.

Table 2: Structural breaks in city population trajectories in Western Europe and Middle East, 800-1800

|  | Minimum breakpoint test statistics | Estimated break date |
|---|---|---|
| *Panel A: Western Europe* | | |
| t-statistics | -30.557 | Around 1200 |
| *Panel B: Arab peninsula* | | |
| t-statistics | -3.585 | Around 1300 |
| *Panel C: Levantine* | | |
| t-statistics | -3.422 | Between 1200 and 1300 |
| *Panel D: North Africa* | | |
| t-statistics | -3.840 | Around 1200 |
| *Panel E: Emirate of Granada* | | |
| t-statistics | -6.918 | Between 1400 and 1500 |

Notes: the table reports Zivot and Andrews (2002) structural break test in the presence of unit root. The test statistics is reported for the year in which the minimum time-specific breakpoint statistics is identified. Possible break points are examined using 15% trimming threshold using simple t-statistics approach to select the optimal number of lags.

## 4    Identification strategy

### 4.1    Setup

Our goal is to examine the contribution of the Shariatic turn and consolidation of Islamic law to the long-term economic development of the Islamic countries for which data on city population exist, that is North Africa and the Middle East compared to Latin Europe consistently. To this end, our aim is to estimate the effect of the consolidation on the trajectories of urban development of the Middle Eastern cities. Therefore, we exploit the density of law schools across Islamic cities before and after the Sharia-based consolidation and systematization of law at the beginning of 13$^{th}$ century as a quasi-experimental source of variation in long-term economic development. Our key identifying assumption that postulates the necessary exclusion restriction states that the consolidation served as a largely unanticipated shock that allows us to further examine the contribution of law schools as the interpreters and guardians of Islamic law to the long-term economic development. Since the foundation of a new law school exhibits the characteristics of time-varying treatment, our identification strategy consists of two parts to facilitate a plausible identification of the long-term effect. First, we compare city size evolution between the Islamic and European cities before and after the consolidation on the pre-break parallel trends assumption between both groups of cities. This may yield a static and more parsimonious estimate of the long-term magnitude of the economic decline of the Islamic cities behind their Western European peers. Our post-Shariatic turn treatment-related variable is interacted with the variable indicating the presence of the university in Western Europe or the density of madrasas in the individual cities. Apart from capturing city-level presence of madrasas, we also construct a richer measure of treatment exposure by interacting the post-Shariatic turn variable with the number of universities or madrasas in the 300 km radius of the city. Second, to estimate dynamic effects of Shariatic turn on the long-term economic development of the Middle East, we construct a more restrictive from treatment. More precisely, we allow for the variation in the timing of newly established law schools after the consolidation of Sharia turn and compare the respective city size trajectories between Islamic and European cities before and after the Shariatic turn in response to the change in the density of law schools. This yields a more dynamic form of two-way fixed-effects triple-differences specification to better unravel the contribution of the consolidation of Islamic law schools to the long-term economic development of the Middle East. Third, to ease the restrictiveness of our estimates, we also relax parallel trend assumption and adopt more distinctive counterfactual approaches such as synthetic control method (Abadie et. al. 2015), which allows us to estimate the average effect of Islamic law schools on city size by estimating the missing counterfactual scenario in the full period after the consolidation of Sharia turn. The counterfactual analysis serves two distinctive purposes. First, estimating the counterfactual city size trajectories may partly reflect the hypothetical trajectory of pre-industrial urban development if the turn towards Sharia never took place and the Islamic cities adopted comparable Western legal reforms. Since the presence of Sharia law is virtually absent from the Latin European city-level donor pool, it is unlikely that the assumption of stable treatment unit value assignment (SUTVA) is violated. And second, the counterfactual analysis may also highlight the extent to which population size of European cities would change in the hypothetical absence of the institutional changes that took place in Latin Europe. If the institutional changes commencing with the legal big-bang in Europe is considered the key form of treatment whilst the Middle Eastern cities may comprise the donor pool, the counterfactual may also highlight the long-term urban development benefits of institutional changes that never evolved in the Middle East during the period of our investigation.

### 4.2  *Static TWFE specification*

Our approach emphasizes potential outcomes framework (Rubin 1974, Robins 1986). In the simplest possible form, let $Y_{i,t}(0,0)$ denote the size of i-th city at time $t$ if the city is unexposed to the Islamic law school at time $t=1$ and $t=2$ to ensure that the city size corresponds with the density path of law schools over time such that $Y_{i,t}(0) = Y_{i,t}(0,0)$ and $Y_{i,t}(1) = Y_{i,t}(0,1)$. Notice that only one of the potential outcomes can be observed for each city which implies that $Y_{i,t} = D_{i,t} \cdot Y_{i,t}(1) - (1 - D_i) \cdot Y_{i,t}(0)$ and ensures that stable city-level treatment assumption holds, and any potential spillover effects can be ruled out. Our causal estimand of interest incorporated into our model is $\lambda_2 = \mathbb{E}[Y_{i,2}(1) - Y_{i,2}(0)|D_i = 1]$ where $D_i$ is the i-th city-level treatment exposure variable.

To unravel the effect of Islamic legal education on pre-industrial long-term economic development by exploiting the Sharia-based consolidation and canonization of Islamic law as a deep institutional change, we estimate a simple triple-differences (DDD) model specification:

$$Y_{i,t} = \eta_0 + \mu_i + \alpha_t + (\text{Law School})_{i,t} \cdot \pi + (1 \cdot [i \in \text{Islamic}] \cdot D_{1,i}) \cdot \delta_1 + (1 \cdot [t > T_0] \cdot D_{2,i}) \cdot \gamma_1 + (1 \cdot [i \in \text{Islamic}] \cdot D_{1,i} \times 1 \cdot [t > T_0] \cdot D_{2,i} \times 1 \cdot [t > T_0] \cdot D_i \times \text{Law School}_{i,t}) \cdot \lambda_1 + \mathbf{X}'_{i,t}\beta + \epsilon_{i,t} \qquad (1)$$

where $Y$ denotes population size of city $i = 1,2,\ldots N$ at time $t = 1,2,\ldots T$, (Law School) is a binary indicator of the university in i-th city, $D_{1,i}$ denotes whether or not each city at time $t$ is under the Muslim rule, $D_{2,i}$ is time-difference variable indicating the period after the consolidation, $\mathbf{X}$ is the vector of the structural covariates of the city-level population size, and $\epsilon$ denotes stochastic disturbances, capturing transitory city shocks. Our key variable of interest is the triple-differenced term $1 \cdot [i \in \text{Islamic}] \cdot D_{1,i} \times 1 \cdot [t > 1200] \cdot D_{2,i} \times 1 \cdot [t > 1200] \cdot D_i \times \text{Law School}_{i,t}$. It denotes the density of law schools in $i$-th city under Muslim rule at time $t$ after Sharia-based consolidation of law in early 13[th] century. Therefore, our key parameter of interest is $\lambda_1$ which captures the contribution of Islamic law schools to city-level population trajectories in the affected cities after the consolidation. Standard errors are simultaneously clustered at city- and year level using finite-sample adjustment of the empirical distribution function through a two-way error component model (Cameron et. al. 2011), and provide a robust inference on the key parameters of interest that do not seem to be tainted by arbitrary heteroskedastic distribution of random error variance or serially correlated stochastic disturbances.

Our estimand of interest captures the average effect of Islamic law schools on the affected cities in the years after the Shariatic turn. The key identification challenge is that $Y_{i,t}(0)$ is never observed for the cities affected by the consolidation of law ($D_i = 1$) whereas our triple-differences coefficient $\lambda_1$ imputes the mean counterfactual for the affected cities by using the change between population size in the Islamic cities and baseline population size in the Western European cities through the pre-conquest parallel trend assumption, intuitively implying that the average city-level population size for Islamic and European cities would have evolved in tandem in the hypothetical absence of the consolidation and canonization and adopting the same legal reforms as Latin Europe. In the potential outcomes framework, the parallel trend assumption implies that $\mathbb{E}[Y_{i,t>12}(0) - Y_{i,t<1200}(0)|D_i = 1, \mathbf{X}_{i,t}] = \mathbb{E}[Y_{i,t>120}(0) - Y_{i,t<12}(0)|D_i = 0, \mathbf{X}_{i,t}]$.

Furthermore, we also assume that the consolidation mimics plausible characteristics of the deep institutional change and as such exhibits no anticipation effect. Not only that this assumption presumes no causal effect of the sacralization of law prior to its realization, but it also implies that city's population trajectory does not vary on any knowledge of the Sunna revival before it was firmly consolidated in the 13[th] century (i.e. Shariatic turn). Under the assumptions of parallel trends and no-anticipation and under independent sampling, $\sqrt{n}(\lambda_1 - \lambda_2) \to \mathbb{N}(0, \sigma^2)$ as $n \to \infty$ and $T$ is fixed which suggests that variance term ($\sigma^2$) is consistently estimable and also allows serially correlated stochastic disturbances at the city level.

### *4.3 Dynamic TWFE specification*

To estimate dynamic effects of Shariatic turn on long-term economic development of the Middle East, the assumption behind the timing of establishing law schools ought to be relaxed to allow for both non-monotonic form of treatment as well as differential timing of law school foundation before and after the Shariatic turn. This allows us to estimate dynamic effect of the turn towards Sharia on long-term development that varies both over time and across cohorts of cities within specific time period. Such setup invokes the characteristics of treatment with staggered adoption (De Chasemartin and D'Haultfoeuille 2020, Roth et. al. 2023).

Let $G_0 = \min\{t: D_{i,t} = 1\}$ be the earliest period in which the law school is founded at the city level. If the city is never treated, then $G_i = \infty$ holds. Under staggered foundation of law schools, city population as a potential outcome can depend on the entire path of the treatment-related assignments. The parallel trend assumption can easily be extended to a more complicated setup with the staggered setting provided that $2 \times 2$ city-year version of assignment holds for all combinations of cities treated at different times. More specifically, parallel trend assumption for the staggered treatment implies that for all $t \neq t'$ and $G_i = g'$, parallel pre-consolidation trends in city size suggest that $\mathbb{E}[Y_{i,t}(\infty) - Y_{i,t'}(\infty)|G_i = g, \mathbf{X}_{i,t}] = \mathbb{E}[Y_{i,t}(\infty) - Y_{i,t'}(\infty)|G_i = g', \mathbf{X}_{i,t}]$ which implies that in the counterfactual where treatment had not occurred, the average city size for all adopting cities would have evolved in parallel with the European control group. Under staggered adoption, parallel trend assumption holds only for years after some units are affected by the underlying treatment (Callaway and Sant'Anna 2021), or only for the groups eventually affected by the treatment but not for the never-treated cities (Sun and Abraham 2021). Staggered establishment of law schools also entails considerable heterogeneity of treatment effect over time (Borusyak et. al. 2021, Goodman-Bacon 2021).

To partially address the heterogeneity of the dynamic effect of Islamic law schools over time, we estimate the following dynamic two-way fixed effects (TWFE) specification:

$$Y_{i,t} = \eta_0 + \mu_i + \alpha_t + \mathbf{Z}'_{i,t}\theta + \left(1 \cdot [i \in \text{Islamic}] \cdot D_{1,i} \times 1 \cdot [t > T_0] \cdot D_{2,i} \times 1 \cdot [t > T_0] \cdot D_i \times \text{Law School}_{i,t} | R_{i,t} = r\right) \cdot \lambda_{1,r} + \mathbf{X}'_{i,t}\beta + \epsilon_{i,t} \qquad (2)$$

where $\mathbf{Z}$ is a coarsened vector of treatment-related variables from eq. (1) used to construct a post-treatment triple-differences variable $1 \cdot [i \in \text{Islamic}] \cdot D_{1,i} \times 1 \cdot [t > 1200] \cdot D_{2,i} \times 1 \cdot [t > 1200] \cdot D_i \times \text{Law School}_{i,t} | R_{i,t} = r$ where we specifically relax the static effect and allow the effect of law schools to vary over time where $r < T$ and $R_{i,t} = t - G_i + 1$ is the time relative to the foundation of new law school across the full sample of treated cities. Without any restriction, dynamic specification per se fails to yield sensible and plausible dynamic effects under the effect heterogeneity across cohorts of cities. Such forbidden comparisons can be avoided if the dynamic effect is estimated for the cohort of cities first affected by the treatment-related variable (Callaway and Sant'Anna 2021) which implies that both parallel trend- and no anticipation assumption can be plausibly identified. By comparing the expected changes in city size for a $g$-th cohort of cities between $g - 1$ and $t$ to that of the control group not-yet-affected at period $t$, the average effect of Islamic law schools on city size is:

$$\lambda_{g,t} = \mathbb{E}[Y_{i,t} - Y_{i,g-1}|G_i = g] - \mathbb{E}[Y_{i,t} - Y_{i,g-1}|G_i = g']$$

which, for any $g' > t$, represents the multi-period analogue of the identification property of the difference-in-differences of population means. It should also be noted that the average treatment effect of Islamic law schools on Islamic cities holds for any plausible comparison group provided that the treatment effect is averaged across similar sets of comparisons. Provided that cohort-specific effect is not tainted by non-zero weight and cross-lag contamination, the average treatment effect of Islamic law schools on Islamic cities is estimated by replacing expectations with their sample analogue which yields:

$$\lambda_{g,t} = \frac{1}{N_g} \sum_{i:G_i=g} [Y_{i,t} - Y_{i,g-1}|\mathbf{X}_{i,t}] - \frac{1}{N_{G_{control}}} \sum_{i:G_i \in G_{control}} [Y_{i,t} - Y_{i,g-1}|\mathbf{X}_{i,t}]$$

where we use never-treated cities denoted as $(G_{control} = \{\infty\})$ as a plausible control group used to assess the contribution of Islamic law schools on long-term development. Since our setup involves a relatively small number of periods and treatment-related city cohorts, we estimate all possible combinations of $(g,t)$ to parse out and unravel the heterogeneity of $\lambda_{g,t}$. One of the chief advantages of the cohort-specific analysis lies in the ability to disentangle the effect of Islamic law schools by cohort of cities establishing new law schools after the Sharia-based consolidation and canonization as well as by the duration of exposure to the newly established law schools. The former allows us to pinpoint specific years in which the effect appears to be strongest whilst the latter allows us to determine somewhat more fully whether the effects tend to be permanent and whether they tend to persist more strongly or weakly over time.

### 4.4 *Estimating the missing counterfactual scenario without parallel trend assumption*

One of the most genuine concerns behind difference-in-differences (DID)-based identification strategy hinges on the validity of parallel trend assumption. If the null hypothesis on the parallel trends is not rejected at plausibly low significance threshold, then conventional DID estimates provide a plausible key parameter on the long-term effect of Islamic law schools on pre-industrial development. By contrast, if the null hypothesis on the parallel trend assumption is rejected, then this would imply that $\mathbb{E}[Y_{i,t}(\infty) - Y_{i,t'}(\infty)|G_i = g, \mathbf{X}_{i,t}] \neq \mathbb{E}[Y_{i,t}(\infty) - Y_{i,t'}(\infty)|G_i = g', \mathbf{X}_{i,t}]$, rendering difference-in-differences estimated effect implausible. To mitigate this concern, we

complement our analysis by estimating the missing counterfactual scenario without necessitating parallel trend assumption through synthetic control method (Abadie et. al. 2015).

Suppose we observe $J+1$ cities in $t = 1, 2, \ldots T$ periods where $t = 800, \ldots 1800$. Suppose a major structural break takes places at time $T_0$ and lasts until $T_{0+1}, \ldots T$ such that $t < T_0$.[14] Let $J$ denote a reservoir of potential control cities unaffected by the Shariatic turn. By relying on potential outcomes framework, suppose that $Y_{i,t}^N$ is the observed realization of the city population size whilst our interest lies in the unobserved counterfactual trajectory of city population that is ought to be estimated, denoted by $Y_{i,t}^I$. By letting $D$ represent a dummy variable indicating the period after the structural break, the outcome realization is as follows:

$$Y_{i,t} = \begin{cases} Y_{i,t}^N & \text{if } D_{i,t<T_0} = 0 \\ Y_{i,t}^I = Y_{i,t}^N + \lambda_1 \cdot D_{i,t} & \text{if } D_{i,t>T_0} = 1 \end{cases} \quad (3)$$

where $Y_{i,t}$ denotes population size in i-th city at time $t$, $Y_{i,t}^N$ is the observed population size in the period before the structural break $(t < T_0)$, and $Y_{i,t}^I$ is the unobserved counterfactual trajectory of population size that has to be estimated, $\lambda_1$ is our key parameter of interest which captures the contribution of the identified breakpoint to city-level population size. Without the loss of generality, the treatment effect of interest after Sharia-based consolidation is:

$$\lambda_{1,t} = Y_{i,t}^I - Y_{i,t}^N = Y_{i,t} - Y_{i,t}^N$$

for $t > T_0$ where our specific gradient is the vector of post-Shariatic city population gaps between the Islamic cities and the control group directly unaffected by conquest, denoted as $\lambda_{1,t} = \{\lambda_{1,t}, \ldots \lambda_{1,T_0}\}$. Since $Y_{i,t}^N$ is unobserved, it must be estimated empirically. To this end, we assume that the counterfactual trajectory of population size is given by the simple latent factor model of the following form:

$$Y_{i,t}^N = \gamma_t + \theta_t \cdot \mathbf{Z}_i + \pi_t \cdot \mu_i + \varepsilon_{i,t}$$

---

[14] In our setup, $T_0$ corresponds to the consolidation and sacralization of law in the 13th century.

where $\gamma_t$ is the unknown common factor with constant factor loading across cities, $\mathbf{Z}_i$ is the vector of full-path pre-Shariatic turn population size variable, $\mu_i$ is $(F \times 1)$ vector of unknown parameters, $\pi_t$ is $(1 \times F)$ vector of unobserved common factors, and $\varepsilon_{i,t}$ denotes stochastic disturbances. The linear factor model provides a plausible extension of difference-in-differences model with fixed effects by letting $Y_{i,t}^N$ depends on multiple unobserved components $\mu_i$ with the corresponding set of coefficients $\pi_t$ that change over over time. Contrary to the difference-in-differences, the linear factor model does not impose parallel trend assumption on the outcome trends for the cities that share similar size values in $\mathbf{Z}_i$. Under the latent factor model, the synthetic control unit reproduces the values of both $\mathbf{Z}_i$ and $\mu_i$, providing reasonably unbiased estimator of the treatment effect for the affected cities.

Let $W = (w_2, \dots w_{J+1})'$ represent a vector of non-zero weights with an additive structure such that $w_j \geq 0$. For each affected city denoted by $j = 2, \dots J+1$, the vector of weights is constructed from the reservoir of unaffected cities such that $\sum_{j=2}^{J+1} w_j = 1$. Each value of $W$ represents a potential synthetic control unit that minimizes the discrepancy in the pre-consolidation values of city-level population between Islamic and European cities. The algorithmic selection of weights is partaken in two stages. In the training stage, the relative importance of the synthetic control in reproducing the values of each of predictor variables for each treated unit is gauged to obtain predictor-specific weights. In the validation stage, each potential candidate for the predictor yields a synthetic control unit, using constrained optimization. Empirically, let $X_1$ be $k \times 1$ vector of pre-conquest characteristics for each affected city subsumed into a vector variable denoted by $\mathbf{X}_1$. Furthermore, let $\mathbf{X}_0$ $k \times J$ matrix containing time variable for the cities directly unaffected by the consolidation. The set of additive weights that falls within the convex hull of the implicit attributes of Islamic cities, $W^*(w_2^*, \dots w_{J+1}^*)$, is chosen to minimize the discrepancy between the affected and unaffected cities through the following normalized Euclidean space norm:

$$\mathbf{W}(\mathbf{V}) = \underset{w \in W}{\mathrm{argmin}} \|\mathbf{X}_{0,t<T_0} - \mathbf{W}'\mathbf{X}_{1,t<T_0}\|_V = \sqrt{\mathbf{X}_{0,t<T_0} - \mathbf{W}'\mathbf{X}_{1,t<T_0}}$$

subject to the constraint that weight sets is both non-negative and equal to unity, where $\mathbf{W}$ denotes the vector of optimal city-specific weights, and $\mathbf{V}$ is the vector of predictor-specific weights used to reproduce the potential synthetic control group that falls within the convex hull of the implicit characteristics of the affected cities prior to the conquest. Under mild regularity conditions, the estimated treatment effect capturing the contribution of Islamic law to long-term development can be written as: $\lambda_{1t>T_0} = Y_{1,t} - \sum_{j=2}^{J+1} w_j^* \cdot Y_{j,t}$ which captures the potential city development trajectory of Islamic cities without the Shariatic turn and assuming the adoption of legal reforms implemented in Latin Europe between 1100 and 1300.

## 5  Results

### 5.1  Static TWFE estimates

#### 5.1.1  Baseline estimates

Table 3 reports the static TWFE triple-differences estimated effect of Islamic and European law schools on pre-industrial economic development. Each specification includes the full set of time-invariant city-specific effects as well as time-varying common technology shocks that simultaneously affect city population trajectories independent of the law schools and madrasas. Columns (1) and (2) exhibit triple-differences estimated long-term effect of law schools in Europe on city-level population. The estimates in column (1) are based on the interaction term between the city-level presence of the law school and post-1100 dummy variable to capture the long-term effects of the legal revolution in European jurisprudence (Berman 1985). Our triple-differences estimates indicate a large and positive effect of law schools on city-level population size. In quantitative terms, triple-differences estimates suggest that the foundation of law schools after 1100 is associated with 74% percent (=exp(.557)) higher city-level population size in the long run, ceteris paribus. The estimated triple-differences coefficient is statistically significant at 1%, respectively (i.e. p-value = 0.000). Columns (2) uses the variation in the number of law schools to capture the exposure to learned law. The estimated triple-differences coefficients confirm somewhat more modest but sizeable effect of the rise of learned law on city-level population size. In particular, our estimates suggest that each additional law schools in 300 km radius after the cultural revolution tends to increase city-level population size by 2.4 percent higher than in the comparable control group without such exposure (i.e. p-value = 0.000). The estimated effect of law schools on long-term pre-industrial development is entirely consistent with Mokyr (2005), Cantoni and Yuchtman (2013), and Schäfer and Wulf (2014) among several others. Each estimated specification confronts both the full vector of covariates from Table 1 as well as a comprehensive set of city-fixed and time-fixed effects to rule out the conjoint influence unobserved heterogeneity on the pre-industrial urban development.

Columns (3) and (4) present a triple-differences estimated effect of Islamic law schools (i.e. madrasas) on long-term development using the variation in the consolidation of the Shariatic turn and city-level density of madrasas. Column (3) uses an isolated city-level density of madrasas as the underlying treatment variable. Contrary to the positive effect of law schools in Europe in column (1), our triple-differences estimates for the Islamic cities uncover evidence of a modest negative effect of the expanding density of madrasas after the consolidation of the Shariatic turn on pre-industrial urban development of the Middle East. In particular, each additional madrasa after the consolidation is associated with 1.3 percent lower city-level population size in the long term in comparison with a comparable control group of European cities without the exposure to madrasas if Islamic territories introduced the entire realm of legal reforms of Latin Europe. The estimated triple-differences coefficient is statistically significant at 5% (i.e. p-value = 0.027). It should be acknowledged that the estimated magnitude of the effect is not very large and is particularly small in comparison with the comparable control group of European cities. Nevertheless, the estimated coefficient is statistically significant at conventional levels. Our estimates do not imply an overarchingly large and negative effect of madrasas on long-term development of Middle Eastern cities. Taking together the magnitude and significance of the triple-differences coefficient, our estimates imply that an increasing exposure to madrasas induced a modest but highly persistent negative effect on the pre-industrial growth, reflecting substantially more dynamic development of learned law and legal institutions in Western Europe as indicated in columns (1) and (2).

Table 3: Long-term effect of universities and madrasas on urban development of Europe and Islamic countries, 800-1800[15]

|  | Europe | | Islamic countries | |
|---|---|---|---|---|
|  | (1) | (2) | (3) | (4) |
| Exposure | Isolated | Network | Isolated | Network |
| Treatment | Binary | Continuous | Continuous | Continuous |
| Panel A: Dependent variable: city population (log) | | | | |
| Post-1100 × Law School | .557*** | | | |
|  | (.104) | | | |
| Post-1100 × # law schools in 300km radius | | .024*** | | |
|  | | (.007) | | |
| Post-1200 Shariatic turn × Madrasa | | | -.013** | |
|  | | | (.007) | |
| Post-1200 Shariatic turn × # madrasas in 300km radius | | | | -.005*** |
|  | | | | (.001) |
| # treatment-control paired observations | 8,723 | 8,723 | 8,723 | 8,723 |
| Structural controls | YES | YES | YES | YES |
| (p-values) | (0.000) | (0.000) | (0.000) | (0.000) |
| City-fixed effects | YES | YES | YES | YES |
| (p-value) | (0.000) | (0.000) | (0.000) | (0.000) |
| Time-fixed effects (p-value) | YES | YES | YES | YES |
|  | (0.000) | (0.000) | (0.000) | (0.000) |

Notes: the table presents the long-term effect of law schools in Europe and the Middle East on long-term pre-industrial economic growth and development using difference-in-difference-in-differences (DDD) specification. Our treatment groups comprise European cities in columns (1) and (2), and Islamic cities in columns (2) and (3). The institutional changes are captured by post-treatment variables capturing the cultural revolution in legal thought (Berman 1985) in Europe, and the consolidation of the Sharia turn in the Middle East. Our third source of variation comprises the density of law schools in both Europe and the Middle East. The dependent variable is the natural log of city-level population size. The standard errors are adjusted for the arbitrary heteroskedasticity and serially correlated stochastic disturbances using finite-sample adjustment of the empirical distribution function through a two-way error component model (Cameron et. al. 20111). Cluster-robust standard errors are reported in the parentheses. Asterisks denote statistically significant coefficients at 10 (*), 5% (**), and 1% (***), respectively.

Furthermore, column (4) uses the variation in the density of madrasas in 300 km radius of each city affected by the Shariatic turn and presents triple-differences estimated coefficient. The evidence reinforces the general tendency of the effect from column (3). In particular, increasing the density of madrasas in the 300km radius of each Islamic city after the fall of Baghdad is associated with 0.5 percent reduction in city-level population (i.e. p-value = 0.000). Although the magnitude of the estimated effect is relatively small, it should not be interpreted as trivial. For instance, expanding the density of madrasas in the 300km radius by one standard deviation (i.e. roughly 23 madrasas) tends to decrease in city-level population size by around 11 percent relative to the European city-level control group, which appears to be both somewhat modestly large.

---

[15] The treatment sample includes the cities from Middle East, North Africa and Spain as only for those regions the data on city size is available. The cities under Islamic rule from Iran, Central Asia and South East Asia are not included in the sample due to data availability constraints.

In the comparative perspective, our results uncover a series of important insights. First, the contribution of law schools in the Middle East and Latin Europe is different after the critical juncture points promulgated a cultural revolution in the legal thought in the latter and profound sacralisation of law in the former, the Shariatic turn. This reflects a different nature of the critical juncture. Whereas the Latin Europe embarked on the trajectory of comprehensive change towards a more scientific treatment of law as a separate discipline and based on individual autonomy, the Middle Eastern societies embarked on the opposite side of path dependence that did not favour a similar change of direction in legal thinking and a similar openness to far reaching legal changes, which promote economic development. The negative coefficient on the interaction term between post-1200 indicator and city-level madrasa density is indicative of this particular disparity. Second, all four specifications include the full set of pre-trends. This implies that if any distinctive systemic or idiosyncratic shock tends to conflate the long-term impact of the Shariatic turn, the coefficient on the interaction term should be null which does not hold. And third, the estimated post-1100 and post-1200 city growth effects do not reflect distinctive historical circumstance, inertia or other time-invariant characteristics since each specification includes the unconstrained set of city- and time-fixed effects, respectively.

Our triple-differences estimates uncover a series of noteworthy insights. Namely, the influence of law schools that gradually spread from Bologna in 1100 to Paris and beyond, was overwhelmingly large and positive and further aided by the cultural and institutional changes induced by the revolution in legal thought. In the Middle East, the Shariatic turn posited a very large institutional change to the city economies which worked in the opposite direction of orienting the law to the Koran and the Sunna tradition. The estimated negative effect of the increasing density of madrasas after the Shariatic turn partly reflects a more vibrant and dynamic evolution of legal institutions in Europe in insurance contracts, freedom of contract and a more rigorous and gradually reinforced jump start of the learned law. These institutional innovations, which were crucially in facilitating the legal thinking supporting low transaction costs and more flexibility in contract and corporate law, had no counterpart in the the Islamic countries covered by this study. Although the estimated magnitudes of the effect of Islamic law schools are relatively small, the effect appears to be highly persistent over time. This implies that independent of the role of Islamic religion, the necessary institutional underpinnings of the early modern capitalism (Van Zanden 2008) simply failed to evolve in a dynamic coherency and comparative institutional advantage witnessed in Western Europe by the early 16$^{th}$ century.

We further elaborate the relationship between law schools and pre-industrial development by unravelling the direction of the hypothetical causality and making use of the test of causation proposed by Granger (1969) and Sims (1980). The general intuition behind such a test is simple. The estimates from Table 4 indicate a strong impetus to the pre-industrial growth of cities provided by the law schools as proxies for the influence of a legal culture that generated an increasing supply of learned legal scholars whereas such development was largely absent in Islamic cites after the Shariatic turn. If this empirical relationship evolves into a dynamic interdependence, then a mutual test of the direction of the effect should emphasize a strong and statistically significant chain of effect running from the foundation of universities to a more vibrant and dynamic economy of medieval cities, and not vice versa. On the contrary, if our estimates from Table 3 are plausible, the absence of any Granger-causal relationship should be perceptible for the Islamic city-level sample.

Table 4 reports the estimated panel-vector autoregressive estimates of the relationship between law schools and city population size together with the Granger causality test. In particular, columns (1) and (2) report the estimates for the European sample whilst columns (3) and (4) report the estimates for the full sample of Islamic cities. Our key parameter of interest is reported in Panel C where the p-values of Wald test associated with the direction of the effect are presented more fully. In particular, if the p-values are sufficiently low the null hypothesis that either variable does not Granger-cause the other variable can be rejected. Our findings reinforce a strong and persistent relationship between the density of law schools and city population size for the European cities, but not for Islamic cities. In panel C, the comparison of p-values on each excluded variable suggests that the null hypothesis that pre-industrial urban development does not lead to systematically higher density of law schools and madrasas cannot be outrightly rejected. By contrast, the null hypothesis that the density of law schools tends to predict systematically higher levels of pre-industrial development can be easily rejected for the European cities either in the full-sample (i.e. p-value = 0.069) or treatment-sample (i.e. p-value = 0.000) specification, but not for Islamic cites. These findings, perhaps somewhat more closely, corroborate the support for our hypotheses, and confirm that dynamic and vibrant development of learned law as a scientific discipline has importantly shaped the contours of pre-industrial growth whilst similar development of legal institutions did not take place in the Middle East after the consolidation of Shariatic turn.

Table 4: Panel vector autoregressive (PVAR) estimated relationship between universities with law schools, Madrasas and city population size across European and Middle Eastern cities, 800-1800

|  | Europe | | Middle East | |
|---|---|---|---|---|
|  | (1) | (2) | (3) | (4) |
| Exposure | Full-Sample | Treatment Sample | Full-Sample | Treatment Sample |
| *Panel A: Dependent variable : city population (log)* | | | | |
| $\log(\text{City Population})_{i,t-1}$ | .211** | .146*** | .055 | .959*** |
|  | (.096) | (.034) | (.172) | (.133) |
| $(\text{Law School})_{i,t-1}$ | .071* | .356*** | | |
|  | (.038) | (.074) | | |
| $(\#\text{Madrasas})_{i,t-1}$ | | | .033 | .041 |
|  | | | (.031) | (.040) |
| *Panel B: Dependent variable: density of law schools / madrasas* | | | | |
| $(\text{Law School})_{i,t-1}$ | .684*** | .076*** | | |
|  | (.107) | (.003) | | |
| $\log(\text{City Population})_{i,t-1}$ | -.014 | .075* | -.621 | -.877 |
|  | (.021) | (.042) | (.549) | (.764) |
| $(\#\text{Madrasas})_{i,t-1}$ | | | .968*** | .964*** |
|  | | | (.040) | (.044) |
| *Panel C: Granger-Wald causality test (i.e. excluded variable)* | | | | |
| Law school (p-value) | 0.069 | 0.000 | | |
| City population (p-value) | 0.498 | 0.265 | 0.258 | 0.251 |
| # Madrasas (p-value) | | | 0.275 | 0.300 |
| # treatment-control paired observations | 7,137 | 5,589 | 7,137 | 1,305 |
| Initial weight matrix | Identity | Identity | Identity | Identity |
| GMM weight matrix | Robust | Robust | Robust | Robust |

Notes: the table reports panel-vector autoregressive estimates of the relationship between the law schools and city-level population sizes for a sample of European and Islamic cities in the period 800-1800. Panel C reports the probability of rejecting the null hypothesis that each excluded variable does not predict the change in the structural equation-level variable. Each specification also includes city-fixed effects which are absorbed into time-demeaning

### 5.1.2  *Cohort-specific TWFE difference-in-differences estimates*

The absence of a more dynamic development of legal institutions in the Middle East confronts two essential issues. First, the aggregate effect potentially masks the heterogeneity in the evolution of the effect over time. It may be informative to learn a more precise magnitude of the effect and its shifts over time which can potentially shed some light as to whether the effect of the absence of more dynamic legal institutions appears to be either relatively constant, temporary or perhaps even permanent after the consolidation of the Shariatic turn, which we consider as a turning point in the pre-industrial growth trajectories of the Middle East. And second, it would be questionable to assume that the Shariatic turn and the subsequent relative stagnation of legal institutions affected all cities in the Middle East equally without any consideration to their size, and the respective importance.

Table 5 reports cohort-specific difference-in-differences estimates of the long-term effect of the absence of dynamic legal innovation after the Shariatic turn using the full treatment sample of Islamic cities, and decomposing the effect across large and small cities.[16] Panel A reports baseline difference-in-differences estimates of the treatment effect after the Shariatic turn. Our covariate-saturated, full-sample specification suggests that the Shariatic turn is associated with 43 percent [=exp(-.578)-1] long-term decline of city-level population size relative to the comparable control group of European cities. After covariate adjustment of our specification for confounding influence, the magnitude of decline appears to be around 62 percent (i.e. p-value = 0.000) for large cities, and 42 percent (i.e. p-value = 0.000) for non-large cities, which implies that large cities such as Baghdad, Cairo and Fez among several others were disproportionately more heavily impacted by the post-1200 stagnation than smaller cities.

---

[16] More specifically, Table 5 reports the estimated $\lambda_{g,t}$ for the full-sample specification as well as for the large-city and small-city specifications

It should be noted that the estimated post-Shariatic turn effect tends to increase noticeably over time. Panel B reports cohort-specific two-way fixed effects estimates of the triple-differences coefficients for the full sample as well as large-city and small-city subsamples. The evidence almost unequivocally suggests that the negative effect tends to increase considerably over time. In our full-sample specification in column (2), the magnitude of the effect tends to increase from -31 percent in the year 1300 to -75 percent in the year 1600 and widens further to -83 percent by the end of 1800. This implies that compared to the sample of European cities, the size of the cities in the Middle East has been considerably and significantly smaller than it would otherwise be in the hypothetical presence of the parallel development of legal institutions in Western Europe and Islamic countries. Since the average size of European cities has rapidly accelerated after the Black Death during the 14$^{th}$ century, the estimated magnitude of the post-1200 effects tends to widen considerably over time. Cohort-specific analysis also reveals that the effect tends to amplify particularly for smaller cities. For instance, TWFE estimates in columns (4) and (6) suggest that the negative effect tends to persist for small and large cities alike whereas the negative estimated effect is particularly large and statistically significant for smaller cities, which appear to be disproportionately more severely affected by the absence of more dynamic and innovative legal institutions than the larger cities. Panel C reports de Chaisemartin and D'Haultfœuille (2020) flexible two-way fixed effects triple-differences estimates where the underlying post-1200 treatment-related variable which avoids the negative weights in the weighted sum of treatment effect and provides a more consistent estimate of the long-term effect of the absence of dynamic legal institutions on pre-industrial development. The evidence suggests that once potentially negative weights in the TWFE-based computation of treatment effect are effectively absorbed, both large and smaller cities tend to be severely affected by the Shariatic turn. The estimated magnitude tends to amplify considerably over time. In particular, our estimates imply that the average population size of small Islamic cities tends to decrease by 31 percent in the year 1300, and up to 83 percent in the year 1800, relative to the comparable control group of European cities. In the large-city specification in column (4), the estimated drop in the average city size is 34 percent in the year 1300 up to 81 percent in our end-of-sample period (i.e. 1800). Figure 8 presents the evolution of the post-1268 effect over time both for the full-sample specification as well as large-city and small-city specifications. Figure 9 separately plots de Chaisemartin and D'Haultfœuille (2020) full-sample estimates of the dynamic effect of $\lambda_1$ and sets the first year of full exposure to Shariatic turn at $t = 0$.

Table 5: Difference-in-differences estimated long-term effect of the Shariatic turn of the Middle East, 800-1800

|  | Full sample | | Large-city sample (Median city-level population size > 90th percentile) | | Small city sample | |
| --- | --- | --- | --- | --- | --- | --- |
|  | (1) | (2) | (3) | (4) | (5) | (6) |
|  | w/o covariates | w covariates | w/o covariates | w covariates | w/o covariates | w covariates |
| Panel A: Baseline difference-in-differences ATET estimate | | | | | | |
| Post-1200 Sharia-based consolidation | -.933*** | -.578*** | -2.089*** | -.986*** | -.912*** | -.559*** |
|  | (.132) | (.133) | (.398) | (.381) | (.143) | (.142) |
| # treatment-control paired observations | 8,701 | 8,701 | 880 | 880 | 8,591 | 8,591 |
| Panel B: Two-way fixed effects (TWFE), cohort-specific difference-in-differences ATET estimate | | | | | | |
| 1300 | -.383*** | -.383*** | -1.319*** | -.222 | -.382*** | .103 |
|  | (.114) | (.114) | (.289) | (.444) | (.114) | (.216) |
| 1400 | -.339** | -.339** | -1.653*** | -.551 | -.338 | .018 |
|  | (.137) | (.137) | (.385) | (.471) | (.138) | (.214) |
| 1500 | -.734*** | -.734*** | -1.981*** | -.450 | -.732*** | -.213 |
|  | (.152) | (.152) | (.441) | (.605) | (.152) | (.248) |
| 1600 | -1.013*** | -1.013*** | -2.247*** | -.654 | -1.012*** | -.614* |
|  | (.171) | (.171) | (.499) | (.784) | (.171) | (.390) |
| 1700 | -1.351*** | -1.351*** | -2.471*** | -.775 | -1.350*** | -.821** |
|  | (.174) | (.174) | (.503) | (.737) | (.174) | (.410) |
| 1800 | -1.776*** | -1.777*** | -2.861*** | -.627 | -1.776*** | -.904* |
|  | (.192) | (.192) | (.504) | (1.114) | (.192) | (.623) |
| Panel C: de Chaisemartin and D'Haultfœuille flexible two-way fixed-effect estimate of heterogenous ATET | | | | | | |
| Baseline $\lambda_1$ | -.436*** | -.312*** | -.641*** | -.417*** | -.388*** | -.287*** |
|  | (.096) | (.102) | (.208) | (.225) | (.109) | (.111) |
| $\lambda_1(t+1)$ | -.392*** | -.307*** | -.975*** | -.841*** | -.289*** | -.217*** |
|  | (.128) | (.131) | (.291) | (.293) | (.142) | (.142) |
| $\lambda_1(t+2)$ | -.787*** | -.608*** | -1.303*** | -1.054*** | -.676*** | -.515*** |
|  | (.137) | (.152) | (.357) | (.346) | (.151) | (.161) |
| $\lambda_1(t+3)$ | -1.066*** | -.769*** | -1.569*** | -1.130*** | -.942*** | -.672*** |
|  | (.165) | (.182) | (.436) | (.426) | (.175) | (.184) |
| $\lambda_1(t+4)$ | -1.404*** | -1.024*** | -1.793*** | -1.251*** | -1.315*** | -.963*** |
|  | (.170) | (.185) | (.451) | (.450) | (.182) | (.199) |
| $\lambda_1(t+5)$ | -1.830*** | -1.123*** | -2.183*** | -1.70*** | -1.631*** | -.957*** |
|  | (.203) | (.243) | (.472) | (.491) | (.212) | (.259) |
| End-of-sample placebo coefficient | -.186 | -.119 | -.572 | -.453 | .133 | .068 |
|  | (.093) | (.094) | (.145) | (.139) | (.114) | (.110) |
| # city-year paired observations | 791 | 791 | 80 | 80 | 711 | 711 |
| # treatment switchers | 113 | 113 | 22 | 22 | 91 | 91 |

Notes: the table presents difference-in-differences (DID) estimated effect of the Sharia-based consolidation of Islamic law on the evolution of city size. Panel A presents baseline difference-in-differences estimate of the average post-1200 treatment effect on the Islamic cities (ATET) both with and without the full set of covariates. Panel B reports cohort-specific two-way fixed effects (TWFE) estimate of the ATET parameter. Panel C reports de Chaisemartin and D'Haultfœuille (2020) flexible difference-in-differences estimates that compare the evolution of city size in the period t-$\mathcal{L}$-1 to period t across $\mathcal{L}$ post-treatment horizon in response to the Shariatic turn to the cities that had been affected by the destruction to the full sample of cities that had never been affected by the Shariatic turn (i.e. non-switchers) which provides plausibly unbiased dynamic and heterogeneous estimate of $\lambda_1$. Standard errors are adjusted for arbitrary heteroskedasticity and within-city serially correlated stochastic disturbances using finite-sample adjustment of the empirical distribution function through two-way error component model. Cluster-robust standard

**Figure 8**: Homogenous treatment effects of the Shariatic turn and absence of Western legal reforms on the development of Islamic cities , 800-1800

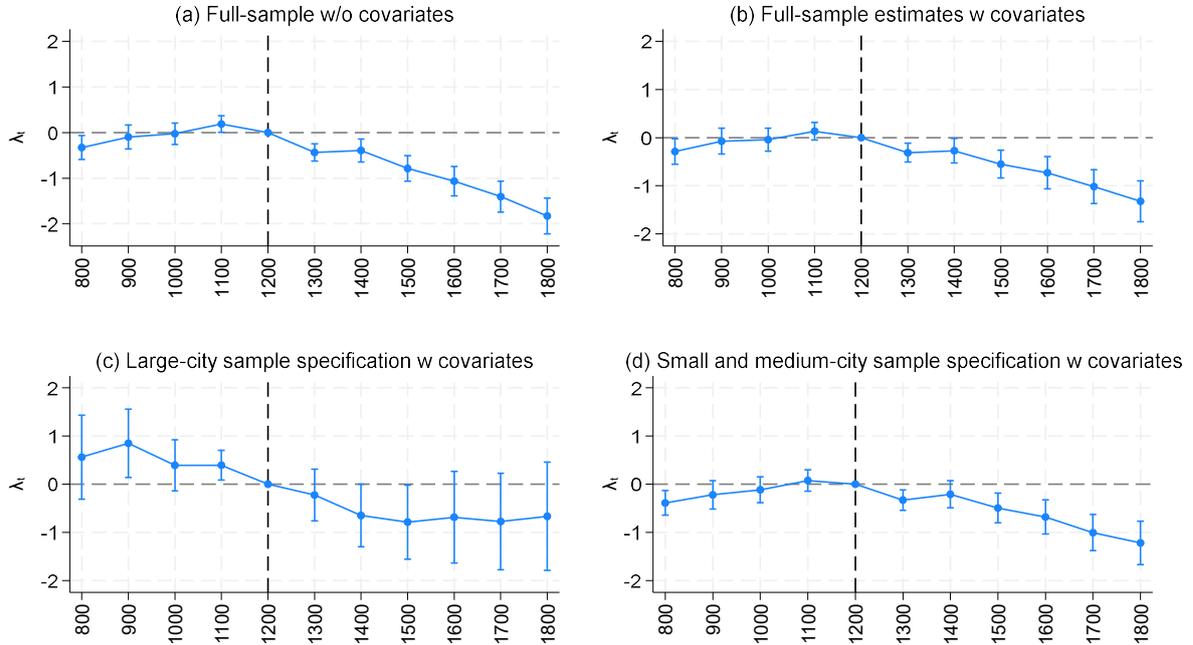

**Figure 9**: Flexible multiple-period difference-in-differences estimated long-term effect of the Shariatic turn on urban development of the Middle East compared to Latin Europe, 800-1800

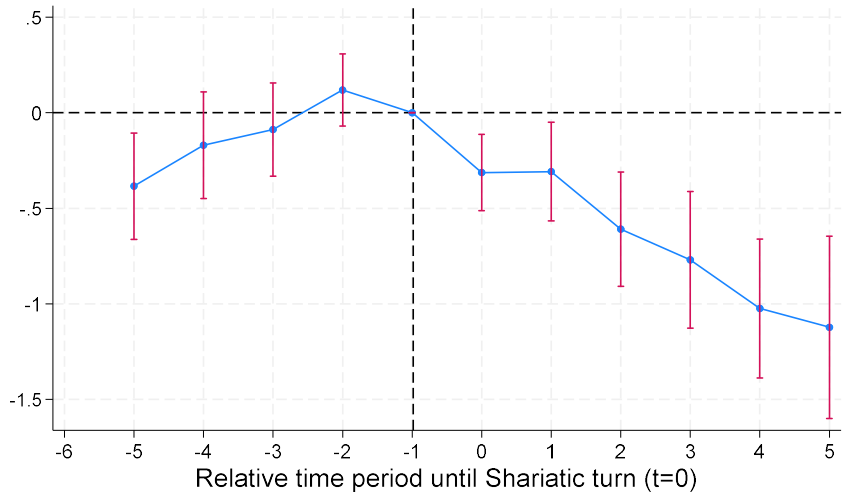

### 5.2 Dynamic TWFE estimates

The empirical evidence on the long-term effect of the absence of growth supporting legal institutions hinges on the notion of the effect homogeneity, implicitly assuming that all Islamic cities are affected at the same time. Nonetheless, the Islamic law schools (i.e. madrasas) that provided legal education were established at different points in time at varying frequencies. The first formal madrasas was established in the city of Fes in the 859 AD. Until 1600, eight additional madrasas were established in the city whilst the number rose to 12 at the beginning of 1800. The city of Constantinople did not establish madrasas until 1400 when two madrasas were established. Until 1600, the number of madrasas increased to 57 and further to 157 by the beginning of 1800. In smaller cities, typically one madrasa was established whilst some cities had no madrasas but many of them were established in the larger neighbouring cities. Varying frequency and differential date of madrasas' establishment imply that the assumption of the homogenous effect of the absence of growth oriented legal institutions can be at least partially brought into question given that the reception of legal education depends both on the number and density of madrasas across and within cities.

We address the potential heterogeneity of the effect by estimating triple-differences model specification with staggered timing of madrasas' foundation from eq. (3) to obtain a dynamic representation of our key parameter of interest, $\lambda_{1,r}$ with varying frequency of the treatment effect. This implies that instead of the binary nature of the treatment, our focus emboldens a continuous treatment variable based on the interaction term between the Islamic cities, density of madrasas and post-Shariatic turn variables. In turn, triple-differences coefficient provides a more compact, yet richer estimate of the dynamic effect of Islamic legal institutions on long-term development in the pre-industrial period based on a more restrictive form of treatment beyond the general city population trend between Islamic and European cities that lies behind comparisons based on the binary form of treatment.

Table 6 reports heterogeneous long-term effects of the Islamic law schools after the Shariatic turn on city-level population size by making use of both standard two-way fixed-effects (TWFE) and inverse probability weighing (IPW) estimators. The latter uses the conditional probability of the treatment assignment using Abadie et. al. (2005) flexible semi-parametric identification approach to identify the long-term effect of interest under regularity conditions. Panel A reports heterogeneity-robust TWFE estimated under staggered timing in the establishment of law schools. Our preferred full-sample estimates in column (1) suggest that the establishment of law schools after the Shariatic turn is associated with 57 percent lower population size in the long run relative to the comparable control group of European cities. IPW-based estimate in column (2) yields a comparable estimate indicating 60 percent drop in the average city size after 1200 in response to the establishment of new law schools. Under staggered establishment of law schools, the estimated magnitude of decline is considerably larger for large cities where the long-run TWFE effect estimate is around 86 percent (i.e. p-value = 0.000) compared to the Western European cities whilst the long-run effect magnitude is somewhat smaller for small cities around 49 percent (i.e. p-value = 0.000), yet statistically significant at 1%. This implies that the establishment of law schools after the Shariatic turn has disproportionately affected the growth trajectories of large cities such as Baghdad and Cairo. The estimated effect tends to increase considerably over time, and is reflective of substantially more dynamic and robust population growth in the control sample of European cities. The increasing magnitude of the negative effect over appears to be robust to model misspecification.

Panel C reports dynamic heterogeneity-robust effect decomposed by the duration of exposure to the establishment of law schools after the Shariatic turn. Such decomposition allows us to capture not only the overall effect of Islamic law schools on long-term city-level development trajectories but also its duration in time after the Shariatic turn relative to the parallel trends in the European cities after the turn. The evidence almost unequivocally confirms the negative impact of the post-1200 law school establishment on city-level population size. Our full-sample estimates confirm a sizeable negative effect of establishing law schools (madrasas) after the Shariatic turn on city-level population size. TWFE estimates in column (1) suggest that post-1200 exposure to the law school establishment of less than 100 years is associated with 32 percent less populous cities compared to Western Europe. The estimated magnitude rises to 62 percent if the exposure length increases to 300 years, and up to 70 percent if the exposure length is 400 years. Consistent with prior evidence in Panel A, the estimated reduction in population size relative to the comparable control group of European cities appears to be somewhat larger for large cities. It should also be noted that due to the inverse treatment propensities, the estimated dynamic effect appears to be somewhat smaller in the initial years of exposure whilst up to the period until 1600, both sets of estimates tend to converge, and arguably confirm the negative effect of post-1200 law schools establishment on the pre-industrial economic development of Islamic cities. Figure 10 plots the evolution of the dynamic heterogeneity-robust treatment effect through the cohort-specific decomposition which uncovers important time-specific similarities and disparities in the effect magnitude. City cohort-level decomposition unveils that the negative effect of establishing law schools is particularly strong in the cohort of cities where madrasas were established between the years 1300 and 1400. The effect appears to be continually negative for the law school establishment cohorts in the years 1500 and 1600 whilst, at the same time, becoming somewhat more temporary until the significance of the effect size tends to disappear for the city-level law school-establishment cohort in 1700.

Table 6: Heterogenous treatment effect of the Shariatic turn on long-term economic development of Islamic countries in Comparison to Western Europe, 800-1800

|  | Full sample | | Large-city sample (Median city-level population size > 90th percentile) | | Small city sample | |
| --- | --- | --- | --- | --- | --- | --- |
|  | (1) | (2) | (3) | (4) | (5) | (6) |
|  | TWFE | IPW | TWFE | IPW | TWFE | IPW |
| Panel A: Heterogeneity-robust estimates under staggered treatment timing | | | | | | |
| Aggregated effect | -.855*** | -.912*** | -1.974*** | -1.304*** | -.687*** | -.815*** |
|  | (.114) | (.108) | (.369) | (.293) | (.114) | (.115) |
| Panel B: Dynamic heterogeneity-robust effect under staggered treatment timing by year after the conquest of Baghdad | | | | | | |
| 1300 | -.366*** | -.418*** | -1.335*** | -.613*** | -.199a | -.365*** |
|  | (.123) | (.099) | (.283) | (.195) | (.125) | (.113) |
| 1400 | -.366*** | -.417*** | -1.601*** | -.898*** | -.176*** | -.341*** |
|  | (.124) | (.114) | (.368) | (.268) | (.129) | (.127) |
| 1500 | -.683*** | -.753*** | -1.929*** | -1.208*** | -.510*** | -.663*** |
|  | (.134) | (.123) | (.407) | (.317) | (.137) | (.132) |
| 1600 | -.958*** | -.998*** | -2.006*** | -1.376*** | -.811*** | -.911*** |
|  | (.145) | (.137) | (.461) | (.382) | (.148) | (.144) |
| 1700 | -1.206*** | -1.247*** | -2.258*** | -1.629*** | -1.067*** | -1.167*** |
|  | (.150) | (.144) | (.467) | (.398) | (.155) | (.154) |
| 1800 | -1.392*** | -1.433*** | -2.612*** | -1.983*** | -1.131*** | -1.231*** |
|  | (.165) | (.174) | (.474) | (.418) | (.168) | (.184) |
| Panel C: Dynamic heterogeneity-robust effect under staggered treatment timing by duration of law school exposure after the Shariatic turn | | | | | | |
| Less than 100 years | -.400*** | -.440*** | -1.208*** | -.574*** | -.309*** | -.409*** |
|  | (.097) | (.081) | (.268) | (.178) | (.104) | (.091) |
| 100 years | -.379*** | -.420*** | -1.472*** | -.834*** | -.258*** | -.360*** |
|  | (.117) | (.106) | (.349) | (.251) | (.124) | (.118) |
| 200 years | -.691*** | -.735*** | -1.834*** | -1.186*** | -.541*** | -.646*** |
|  | (.131) | (.123) | (.396) | (.306) | (.136) | (.134) |
| 300 years | -.987*** | -1.054*** | -2.158*** | -1.458*** | -.808*** | -.954*** |
|  | (.153) | (.145) | (.457) | (.389) | (.158) | (.155) |
| 400 years | -1.231*** | -1.313*** | -2.428*** | -1.734*** | -1.068*** | -1.233*** |
|  | (.160) | (.160) | (.480) | (.420) | (.166) | (.173) |
| 500 years | -1.745*** | -1.819*** | -2.893*** | -2.174*** | -1.425*** | -1.592*** |
|  | (.196) | (.203) | (.494) | (.449) | (.204) | (.221) |
| # never-treated obs | 7,051 | 7,051 | 594 | 594 | 6,457 | 6,457 |
| # city-year paired obs | 8,723 | 8,723 | 880 | 880 | 7,843 | 7,843 |

Notes: the table presents difference-in-differences-differences (DDD) estimated effect of the establishment of law schools after post-1200 Shariatic turn on the evolution of city size for the full treatment sample of Islamic cities using the European city-level control sample. Panel A presents heterogeneity-robust estimate of the overall effect post-consolidation law school establishment on the city-level population size (ATET) and includes the full set of covariates. Panel B reports time-specific estimate of the ATET parameter assuming staggered date of law schools' establishment. Panel C reports dynamic heterogeneity- robust effect of the post-Shariatic turn establishment of law schools decomposed by time-specific duration of exposure. Large cities are defined as those with the mean city-level population size in the 90[th] percentile of the entire pre- and post-Shariatic turn distribution. Standard errors are adjusted for arbitrary heteroskedasticity and within-city serially correlated stochastic disturbances using finite-sample adjustment of the empirical distribution function through two-way error component model. Cluster-robust standard errors are reported in the parentheses. Asterisks denote statistically significant ATET parameters at 10% (*), 5% (**), and 1% (***), respectively.

**Figure 10**: Heterogeneity-robust difference-in-differences estimated effect of the Shariatic turn on economic development in Islamic countries, 800-1800

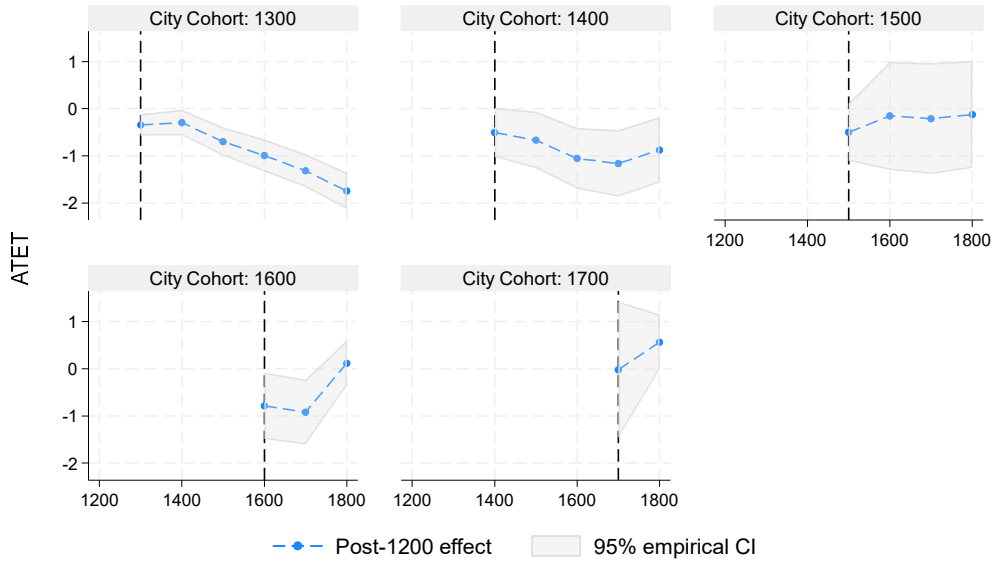

One notable advantage of the inverse propensity weighing approach to estimate the post-Shariatic turn treatment effect associated with the establishment of law schools is the ability to leverage post-consolidation effect size against the pre-Shariatic turn effect trends by comparing the lags and leads of our continuous treatment. The intuition behind such approach is simple. If the law schools establishment prior to the Shariatic turn exhibits a non-zero and possibly large effect prior to the Shariatic turn, then it is likely that our estimates capture some discernible pre-Shariatic turn trend that cannot be associated with the law school establishment after the Shariatic turn. On the contrary, if the null hypothesis on pre-Shariatic turn continuous treatment effect of law school establishment cannot be rejected at conventional threshold, then it is likely that our analysis provides some evidence of the significant effect of post-Shariatic turn law school establishment on city-level population size. Figure 11 reports IPW-estimated pre-Shariatic turn and post-Shariatic turn treatment effect of the increasing number of law schools' establishment on the city-level pre-industrial development for each law school establishment cohort. The evidence generally suggests that the null hypothesis on pre-Shariatic turn effect of law schools can be generally rejected. Not a single estimated specification yields plausible evidence of the non-rejection of the null hypothesis which implies that pre-Shariatic turn trends do not pose a strong concern for the validity of our identification strategy. Consistent with the dynamic heterogeneity-robust estimates in Table 7, the evidence uncovers important insights and effect disparities. For instance, the negative effect of the increasing number of law schools after 1200 appears to be particularly large, negative and permanent for the 1300 and 1400 adoption cohorts. That is, the largest estimated relative reduction in the average size of cities is perceptible in the first 150 years after the consolidation of the Shariatic turn. The estimated dynamic coefficient tends to increase sporadically over time and indicates an important source of population development slowdown compared to the European cities' control sample. For the adoption cohorts 1500 and 1600, the estimated effect becomes noticeable weaker, and readily implies that the absence of more dynamic legal institutions that characterized Islamic law schools after the Shariatic turn is an important source of long-term differences in pre-industrial development between Islamic and Western European cities. The empirical 95% confidence intervals behind the effect sizes are shaded within the figure.

Figure 11: Heterogeneity-robust IPW difference-in-differences estimated effect of the Shariatic turn on economic development of Islamic countries in comparison to Western Europe with pre-treatment trends, 800-1800

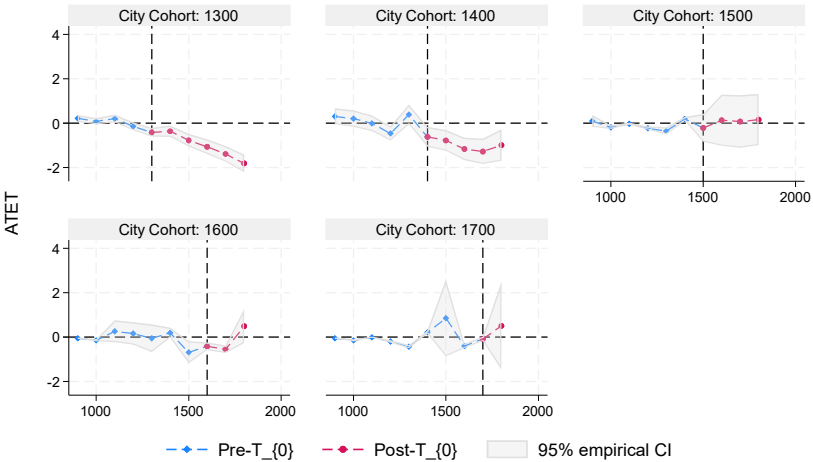

We further address the robustness of our dynamic estimates to the violation of parallel trend assumption. In particular, we assess the plausibility of the assumption by reconstructing our triple-differences design into the event study analysis using post-1200 establishment of law schools as the underlying treatment using dynamic heterogeneity-robust estimates as our starting point. If the validity of the parallel trend assumption is parametrically questionable, the event-study analysis should yield the set of dynamic effects of post-Shariatic turn law schools' establishment that could be seldom compatible with heterogeneity-robust estimates.

Table 7 reports reconsidered heterogeneity-robust estimates of the dynamic effect of post-1200 law schools establishment using two varieties of large- and small-city sample specifications together with the full-sample counterpart. Estimates in Panel A report Callaway and Sant'Anna (2021) multiple-period doubly-robust estimates of the long-term effect of post-Shariatic turn law schools establishment on city-level population size. Our ATET estimates confirm our baseline result and suggest that post-1200 establishment of law schools in Islamic cities is associated with around 59 percent reduction in the average city size compared to the plausible European city-level counterfactual. Decomposing the effect by time-specific establishment cohorts implies that a strong negative effect is perceptible for the years 1300 and 1400 whilst it tends to gradually dissipate up to the end-of-sample adoption cohort. Columns (3) and (4) present the restricted large-city model estimates and indicate a considerably large negative effect which exceeds its full-sample counterpart in columns (1) and (2) by 42%, indicating a substantially more pronounced negative impact for the large cities. For the large cities, the negative effect appears to be statistically significant at conventional thresholds for the adoption cohorts from the years 1300 to 1500, which is entirely consistent with our heterogeneity-robust cohort-specific estimates in Figure 6. Columns (5) and (6) use small-city specifications, confirming a sizeable negative effect of the post-1200 establishment of law schools which appears to be around 38% smaller compared to the large-city ATET estimate in columns (3) and (4). The negative and significant ATET estimate is perceptible in the city-level adoption cohorts between 1300 and 1400 when the effect tends to peak. Regardless of the size-related specification restrictions, Panel C presents a time-specific decomposition of the effect, and confirms that altogether, the post-1200 effect of the law schools establishment on city-level population size tends to be large and negative. Against this backdrop, Panel D presents Butts and Gardner (2021) two-stage triple-differences estimates where Islamic and European cities are compared by removing city and year-fixed effects to deliver a more robust heterogeneity-adjusted treatment effects. It also presents Borusyak et. al. (2021) imputation-based treatment-effect estimates without strong restrictions on effect homogeneity. The evidence largely suggests that the re-estimated post-1200 effect of law schools establishment is negative, statistically significant and robust against model misspecification. It also appears to be particularly stronger and higher for large cities compared to the small cities. Figure 12 plots pre- and post-consolidation coefficients on the madrasas establishment for full-sample specification as well as large and small cities. Given that the null hypothesis on pre-Shariatic turn continuous trends of madrasas establishment cannot be rejected across the full spectrum of estimated specification, it is unlikely that pre-Shariatic turn trends are the most likely shaping force behind the post-1200 effects. The estimated magnitudes of the post-1200 madrasas effect appear to be relatively large and permanent. The null hypothesis on the equality of post- and pre-consolidation law event-style coefficients is consecutively rejected (i.e. p-value = 0.000)

Table 7: Heterogenous treatment effect of the Shariatic turn on long-term economic development of Islamic countries in comparison to Western Europe under alternative estimators, 800-1800

|  | Full sample | | Large-city sample (Median city-level population size > 90th percentile) | | Small city sample | |
|---|---|---|---|---|---|---|
|  | (1) | (2) | (3) | (4) | (5) | (6) |
|  | TWFO | IPW | TWFO | IPW | TWFO | IPW |
| Panel A: Callway and Sant'Anna (2021) heterogeneity-robust, multiple-period treatment effect estimate | | | | | | |
|  | Sant'Anna and Zhao (2020) doubly robust estimator | Abadie (2005) IPW DiD estimator | Sant'Anna and Zhao (2020) doubly robust estimator | Abadie (2005) IPW DiD estimator | Sant'Anna and Zhao (2020) doubly robust estimator | Abadie (2005) IPW DiD estimator |
| Aggregate ATET | -.912*** | -.912*** | -1.304*** | -1.303*** | -.815*** | -.815*** |
|  | (.108) | (.108) | (.293) | (.293) | (.115) | (.116) |
| Panel B: Callway and Sant'Anna (2021) heterogeneity-robust, cohort-specific treatment effect estimate | | | | | | |
| 1300 | -.973*** | -.973*** | -1.386*** | -1.386*** | -.847*** | -.847*** |
|  | (.128) | (.128) | (.311) | (.311) | (.139) | (.140) |
| 1400 | -.969*** | -.969*** | -.273** | -.273** | -1.002*** | -1.002*** |
|  | (.218) | (.219) | (.121) | (.121) | (.226) | (.226) |
| 1500 | .033 | .033 | -.662*** | -.662*** | .216 | .216 |
|  | (.476) | (.476) | (.127) | (.126) | (.517) | (.517) |
| 1600 | -.163 | -.163 | -.068 | -.068 | -.164 | -.164 |
|  | (.142) | (.142) | (.093) | (.093) | (.155) | (.157) |
| 1700 | .206 | .206 |  |  | .177 | .177 |
|  | (.490) | (.491) |  |  | (.490) | (.492) |
| Panel C: Callway and Sant'Anna (2021) heterogeneity-robust, time-specific treatment effect estimate | | | | | | |
| 1300 | -.418*** | -.418*** | -.613*** | -.613*** | -.365*** | -.365*** |
|  | (.099) | (.100) | (.195) | (.195) | (.113) | (.113) |
| 1400 | -.417*** | -.417*** | -.898*** | -.898*** | -.341*** | -.341*** |
|  | (.114) | (.115) | (.268) | (.268) | (.127) | (.127) |
| 1500 | -.753*** | -.753*** | -1.208*** | -1.208*** | -.663*** | -.663*** |
|  | (.123) | (.123) | (.317) | (.317) | (.132) | (.137) |
| 1600 | -.998*** | -.998*** | -1.376*** | -1.376*** | -.911*** | -.911*** |
|  | (.137) | (.137) | (.383) | (.382) | (.144) | (.145) |
| 1700 | -1.247*** | -1.247*** | -1.629*** | -1.629*** | -1.167*** | -1.167*** |
| | (.144) | (.145) | (.398) | (.398) | (.154) | (.154) |
| 1800 | -1.433*** | -1.433*** | -1.983*** | -1.983*** | -1.232*** | -1.231*** |
| | (.174) | (.175) | (.418) | (.418) | (.184) | (.184) |
| Panel D: Gardner (2021) and Borusyak et. al. (2021) dynamic heterogenous treatment effect | | | | | | |
| Overall effect | -.855*** | -.614*** | -1.974*** | -1.902** | -.680*** | -.509*** |
| | (.114) | (.118) | (.360) | (.443) | (.114) | (.118) |
| Empirical 95% confidence interval | (-1.079, -.631) | (-.847, -.382) | (-2.680, -1.268) | (-1.961, -.223) | (-.911, -.463) | (-.742, -.276) |
| # paired obs | 8,723 | 8,723 | 880 | 880 | 7,843 | 7,843 |

Notes: the table reports heterogeneity-robust triple-differences estimates of the long-term effect of the post-traditionalist consolidation of Islamic law schools establishment on city-level population size. Panel A reports Callaway and Sant'Anna (2021) multiple-period treatment effects using both Sant'Anna and Zhao (2020) doubly robust estimator under regularity conditions to achieve semi-parametric estimate efficiency and Abadie (2005) inverse probability weighing estimator based on propensity score framework. Establishment-cohort and time-specific heterogeneity-robust estimates. Panel D reports Butts and Gardner (2021) two-stage triple-differences and city-time demeaned estimates together with Borusyak et. al. (2021) imputation-based treatment effects based on staggered timing of madrasas establishment. Large cities are defined as those with the mean city-level population size in the 90[th] percentile of the entire pre- and post-1268 distribution. Standard errors are adjusted for arbitrary heteroskedasticity and within-city serially correlated stochastic disturbances using finite-sample adjustment of the empirical distribution function through two-way error component model. Cluster-robust standard errors are reported in the parentheses. Asterisks denote statistically significant ATET parameters at 10% (*), 5% (**), and 1% (***), respectively.

**Figure 12**: Event-study pre- and post-Shariatic turn dynamic effects of the establishment of madrasas on long-term pre-industrial development

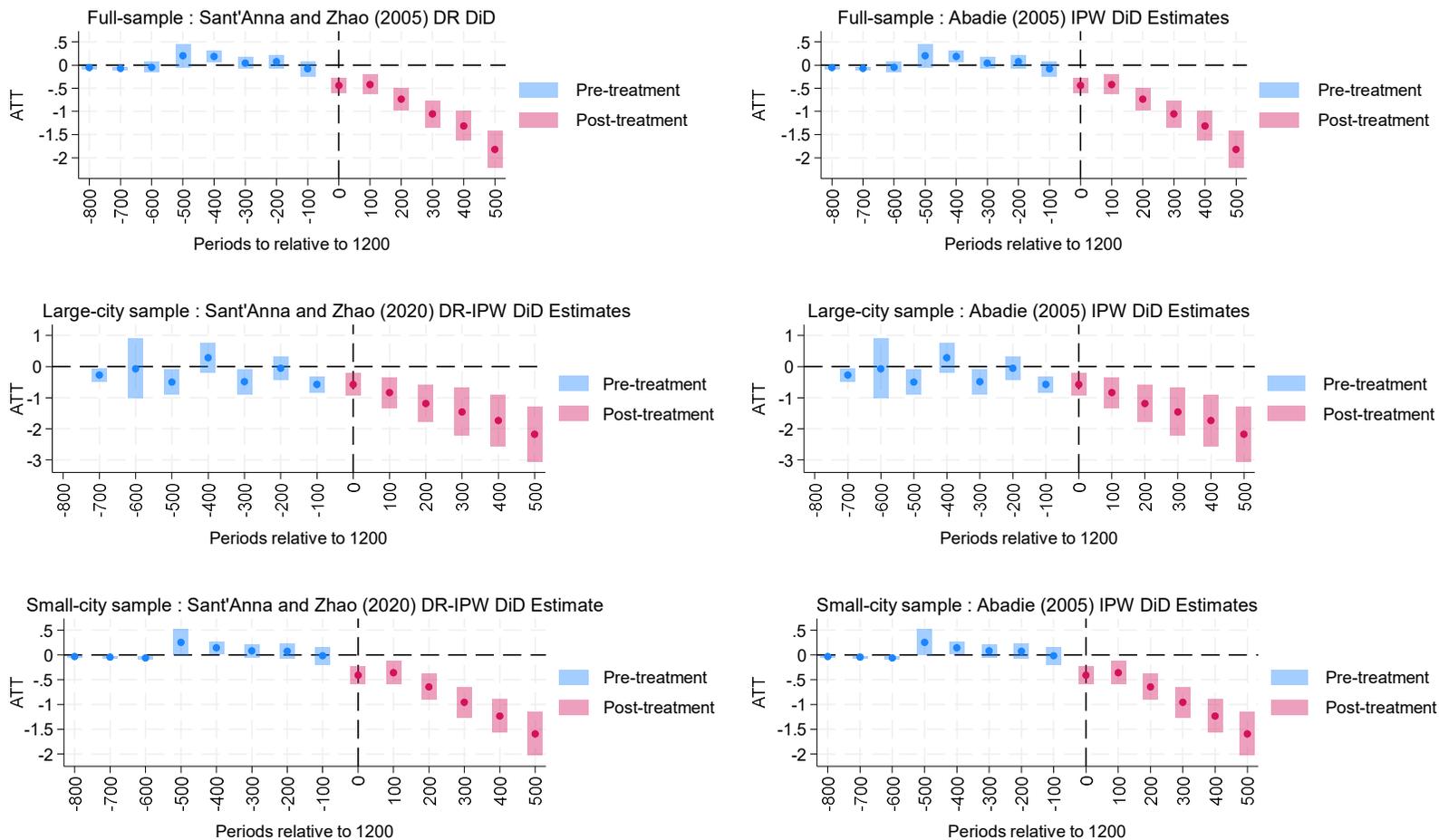

## 5.3 Synthetic control estimates

Triple-differences estimates of the long-term effect of post-1200 establishment of law schools indicate Islamic law schools after the Sharia-based consolidation around 1200 as one of the potential source of Middle East's comparative economic decline behind Western Europe. The validity of our triple-differences estimates hinges on the parallel trend assumption. It should be noted that not all Islamic cities were created equal. Some of them such as Baghdad and Cairo evolved into worldwide centres of learning and commerce and attained an unparalleled level of prosperity. This implies that considering all Islamic cities as a single treatment group may lead to the compressed comparison of Islamic cities with Western Europe that may not be fully informative. At the same time, cities such as Baghdad and Cairo were large and spectacular metropolises at the height of the Islamic golden age which renders a parallel trend assumption somewhat questionable. In the light of the notable effect heterogeneity and failure of the affected cities attributes outside the Western European common support range, these questions necessitate the application of the synthetic control-related methods to estimate and better understanding the missing counterfactual scenario.

### 5.3.1 Interactive synthetic control estimates

Table 8 reports the estimated counterfactual population size trajectories of two of the largest cities in the late medieval Islamic empire (i.e. Baghdad and Cairo) based on interactive-fixed effects algorithm of Xu (2017) generalized version of the synthetic control method. The key advantage of the interactive fixed-effects algorithm is the augmented factor model (Bai 2009) where factor loadings are obtained for each city through the minimization of the mean squared error term for the predicted city size in the pre-destruction period. Given the scarcity of any rule for the number of factors included in the model, we employ a cross-validation scheme before the post-Shariatic turn city size effects are estimated. The cross-validation consists of three steps. First, pre-specified number of factors is identified by absorbing the number from the donor pool of Western European cities through an OLS regression using the rest of pre-Shariatic turn data to find optimal factor loadings. Second, cross-validation loop across the entire spectrum of pre-Shariatic turn years is run in two stages. In the first stage, city sizes for Baghdad and Cairo are predicted once mean square prediction error is calculated by iteratively repeating the procedure for a different number of factors until the distribution of MSPE is built. In the second stage, RMSE-minimizing number of factors is selected. The advantage of this approach is that by incorporating the interactive fixed effects variables into the cross-validation scheme, pre-1200 imbalance in city size between Baghdad, Cairo and their Western European control groups that share similar population size attributes can be almost entirely minimized. In addition, we estimate the city size effect associated with post-1200 foundation of madrasas both for the overall post-destruction period as well as for each post-destruction year separately.

Figure 13 reports interactive fixed-effects estimated population size gaps for Baghdad and Cairo associated with the Shariatic turn in Islamic law. Panel A reports the overall effect for both cites. Our synthetic control estimates invariably indicate a high degree of population decline behind Western European cities. The overall estimated decline of Baghdad after 1200 indicated by our estimates is around 94 percent relative to the comparable latent group of Western European cities. The magnitude of the decline appears to increase substantially over time. Our interactive synthetic control estimates tend to amplify from -36 percent (i.e. p-value = 0.000) in the year 1300 to -96 percent (i.e. p-value = 0.000) by 1600 and remain stable thereafter. The estimated magnitude of post-destruction decline for Cairo is somewhat smaller. Overall, our interactive synthetic control estimates indicate around 50 percent reduction in Cairo's city size after 1200 relative to the Western European control group. The estimated magnitude is in the range between -31 percent in 1300 to -53 in 1600, respectively. Figure X presents the estimated magnitude of the city size decline for both cities graphically alongside 95% empirical confidence intervals.

Table 8: Interactive synthetic control estimated effect of the Shariatic turn in Islamic Law on the city size of Baghdad and Cairo

|  |  | Baghdad | Cairo-Fustat |
|---|---|---|---|
|  |  | (1) | (2) |
|  |  | Interactive fixed-effects algorithm | Interactive fixed-effects algorithm |
| Panel A: Overall ATT estimate |  |  |  |
|  | $\lambda_1$ | -2.494*** | -.695*** |
|  |  | (.073) | (.034) |
| Empirical 95% confidence intervals |  | (-2.611, -2.339) | (-.764, -.641) |
| Panel B: Estimated ATT by year |  |  |  |
|  | $\lambda_{1,1300}$ | -.451*** | -.378*** |
|  |  | (.145) | (.067) |
|  | $\lambda_{1,1400}$ | -.674*** | -.005 |
|  |  | (.142) | (.067) |
|  | $\lambda_{1,1500}$ | -1.508*** | -.723*** |

|  |  |  |
|---|---|---|
|  | (.181) | (.065) |
| $\lambda_{1,1600}$ | -3.286*** | -.772*** |
|  | (.156) | (.005) |
| $\lambda_{1,1700}$ | -4.407*** | -.554*** |
|  | (.165) | (.068) |
| $\lambda_{1,1800}$ | -4.640*** | -1.745*** |
|  | (.131) | (.067) |

*Notes*: the table presents the average city size effect (ATT) of the Sharia-based consolidation in Islamic law on the population size trajectories of Baghdad and Cairo for the period 800-1800 based on the comparison with a donor pool of 621 European cities. Panel A presents full-treatment sample, overall estimates whereas Panel B decomposes the estimated effect by year. Standard errors are adjusted for serially correlated stochastic disturbances both across and within cities using 1,000 bootstrap replications through random sampling with replacement to correct the estimated ATT for the spatial and temporal idiosyncrasies. Standard errors are reported in the parentheses. Asterisks denote statistically significant ATT coefficients at 10% (*), 5% (**), and 1% (***), respectively.

**Figure 13**: Long-term effect of the Shariatic turn in Islamic law on pre-industrial economic development of Baghdad and Cairo in comparison with western Europe, 800-1800

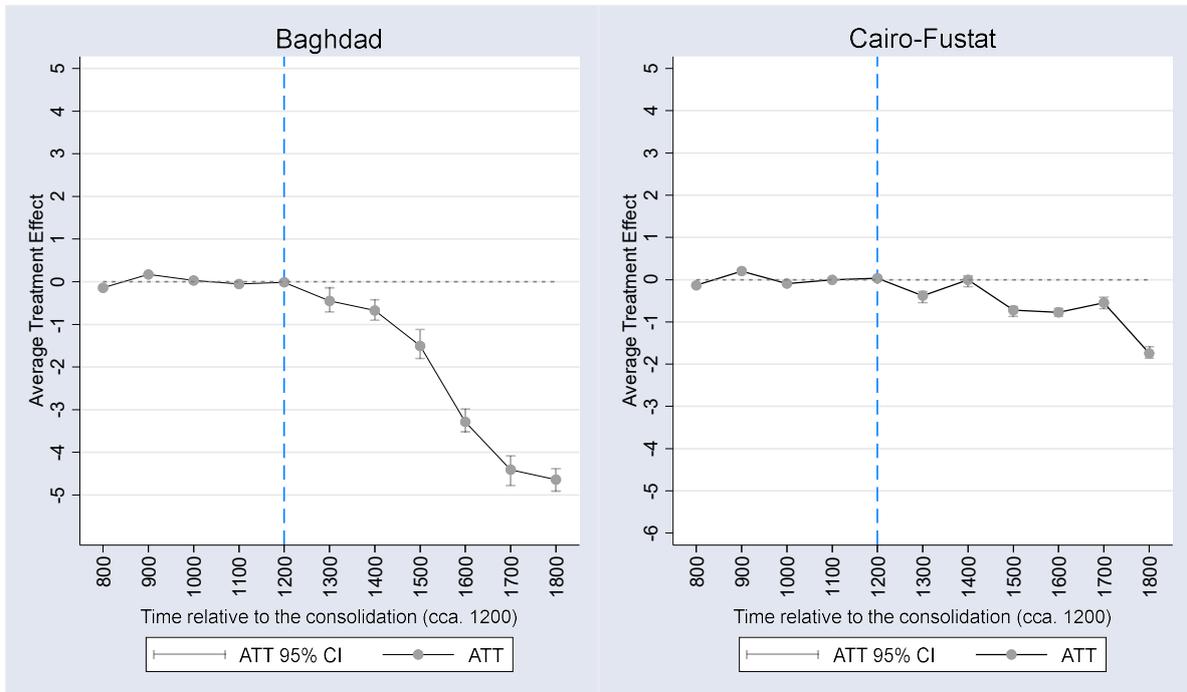

### 5.3.2 *Classical synthetic control estimates*

The population size attributes of other Islamic cities intrinsically fall within the common support range of the Western European cities which implies that pre-Shariatic turn synthetic matching is possible and that the counterfactual trajectory of population dynamics can be estimated upfront using classical synthetic control estimators. At the same time, since the size of cities is comparable in both absolute and relative terms with the control sample of Western European cities, our dependent variable is untransformed and, unlike in the analysis of Baghdad and Cairo, denotes population size without natural log transformations. Our analysis is more or less parsimonious and is conducted on a sample of large and smaller cities to better understand the magnitude of the post-destruction decline.[17]

*5.3.2.1 Large city-level evidence*

Table 9 lays out the estimated synthetic counterfactual population dynamics for four selected cities in the Middle East (i.e. Damascus, Fez, Raqq'a and Damietta). The evidence from the synthetic control analysis of the post-Shariatic turn decline arguably indicates large-scale reduction of city size afterwards. Our estimated average city size effect is in the range between -76,500 inhabitants in Damietta and -133,211 inhabitants in Fez. Across all four respective cities, the estimated overall effect of the Shariatic turn is statistically significant at 1% and is evaluated at the end-of-sample period through non-parametric permutation of the Shariatic turn to the entire donor pool of European cities. It should also be noted that the depth of the decline tends to amplify substantially over time. For instance, the approximate population outflow in Damascus after 1268 rises from 24,700 inhabitants to more than 124,000 inhabitants by the end of 1600. A similar increasing magnitude of the effect is perceptible for other cities reported in the table. For instance, by 1600, our estimates show that Fez and Raqq'a tend to have around 135,390 and 140,710 fewer inhabitants than their comparable European city-level synthetic control groups with similar population size and dynamics unaffected by the Shariatic turn.

**Table 9**: Synthetic control estimated long-term effect of the Shariatic turn in Islamic law on long-term evolution of city size in the Middle East, 800-1800

|  | Large-City Treatment Samples | | | |
|---|---|---|---|---|
|  | (1) | (2) | (3) | (4) |
|  | Damascus | Fez | Raqq'a | Damietta |
| ATET | -95.5*** | -133.21*** | -104.65*** | -76.50*** |
|  | (.0000) | (.0000) | (.0000) | (16.836) |
| Treatment effect by year: |  |  |  |  |
| 1300 | -24.70 | -79.66 | -12.63 | -5.15 |
| 1400 | -10.56 | -61.83 | -30.42 | 0.280 |
| 1500 | -33.68 | -53.39 | -70.43 | -29.820 |
| 1600 | -124.08 | -135.39 | -140.71 | -138.52 |
| 1700 | -152.82 | -216.48 | -156.51 | -124.92 |
| 1800 | -227.14 | -252.54 | -217.09 | -160.87 |

Notes: the table reports synthetic control estimated difference between the observed city size and the synthetic counterfactual of the European cities unaffected by the Shariatic turn for the entire period after 1200. Standard errors are adjusted for arbitrary heteroskedasticity and within-city serially correlated stochastic disturbances using finite-sample adjustment of the empirical distribution function through two-way error component model. Cluster-robust standard errors are reported in the parentheses. Time-specific treatment effect of the Shariatic turn in Islamic law are obtained based on the standard treatment permutation

---

[17] Additional analyses for a larger group of cities are available upon request

proposed by Abadie et. al. (2010) by obtaining the distribution of non-parametric p-values and the associated effect size magnitudes. Asterisks denote statistically significant ATET coefficients at 10 (%), 5% (**), and 1% (***), respectively.

It should be stressed that our synthetic control analysis uses a localized full outcome-path approach in the training and validation stage to find optimal European city-level weights to construct the synthetic counterfactual population size trajectories. Since full outcome path-based optimization absorbs the entire variation in the pre-destruction outcome into the **X** vector, our analysis provides an excellent quality of the pre-Shariatic turn fit between the Islamic and European cities with very little pre-1300 outcome imbalance. Figure 14 presents the a series of the estimated synthetic counterfactual city size trajectories for the selected cities. The comparison of gaps after the Shariatic turn reveals some insights into the gradually more expansive gap behind Western Europe. For instance, the observed estimated population size of Damascus in the year 1600 is around 70,000 inhabitants. Our synthetic control estimates imply that in the hypothetical absence of the Shariatic turn, the population size of Damascus would be approximately 194,000 inhabitants. In terms of observed outcomes, the synthetic counterfactual size implies that Damascus would be the fourth largest city worldwide after Paris, Naples and London. Similarly, large and incremental population size gains are evident for the other cities. For example, the city of Fez' population size in the year 1600 is around 50,000 inhabitants. Our counterfactual estimates suggest that without the Shariatic turn but with the legal innovations in Latin Europe around 1200, the approximate population size would likely reach around 185,000 inhabitants which would place Fez in the size rank between Venice (151,000 inhabitants) and London (200,000 inhabitants). Without the loss of generality, it appears that the Shariatic turn in comparison to the legal revolution in Western Europe law appears to be have induced a permanent effect on the evolution of Middle Eastern cities.

**Figure 14**: Long-term effect of the Shariatic turn on city-level population size trajectories of large cities across the Middle East and North Africa, 800-1800

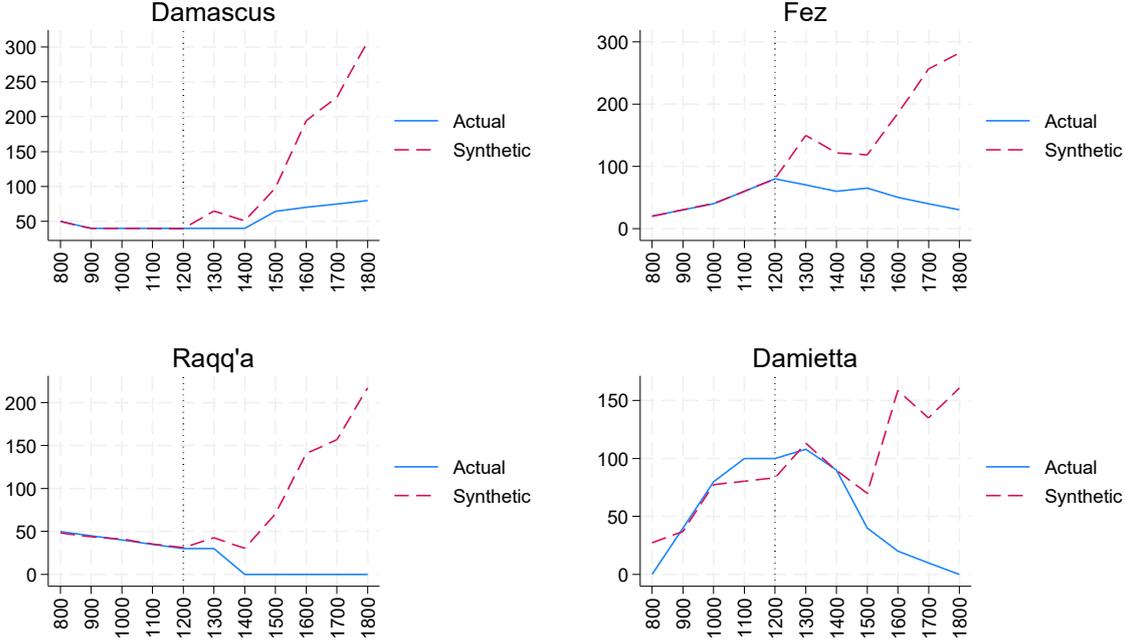

An important question pertains to the composition of synthetic control groups for the Middle Eastern cities. City-level synthetic control groups reveal most salient characteristics that best track the population size trajectories of the Middle Eastern cities before the Shariatic turn. Figure 15 exhibits a detailed composition of synthetic control groups for the four considered cities. For instance, pre-Shariatic turn population size trajectory of Damascus is best reproduced by a convex combination of the implicit characteristics of Naples (55%), Rome (18%), Cordoba (15%), Paris (7%), Murcia (8%), and Verona (<1%). In varying weight-specific proportions, these cities best reproduce the Damascus's trajectory of population size prior to the Shariatic turn. In addition, the population size trajectory of Fez prior to the siege is best synthesized by a convex combination of the size attributes of Paris (41%), Venice (24%), Sevilla (15%), Cordoba (8%), as well as Milan and Florence with a minor weight share. Somewhat similar to Damascus, Raqq'a synthetic control group consists of Naples (43%), Merida (27%), Rome (15%) and Cordoba (15%) whereas Damietta's synthetic control group loads strongly on Sevilla (80%) and, to a lesser extent, on Paris (14%) and Venice (7%).

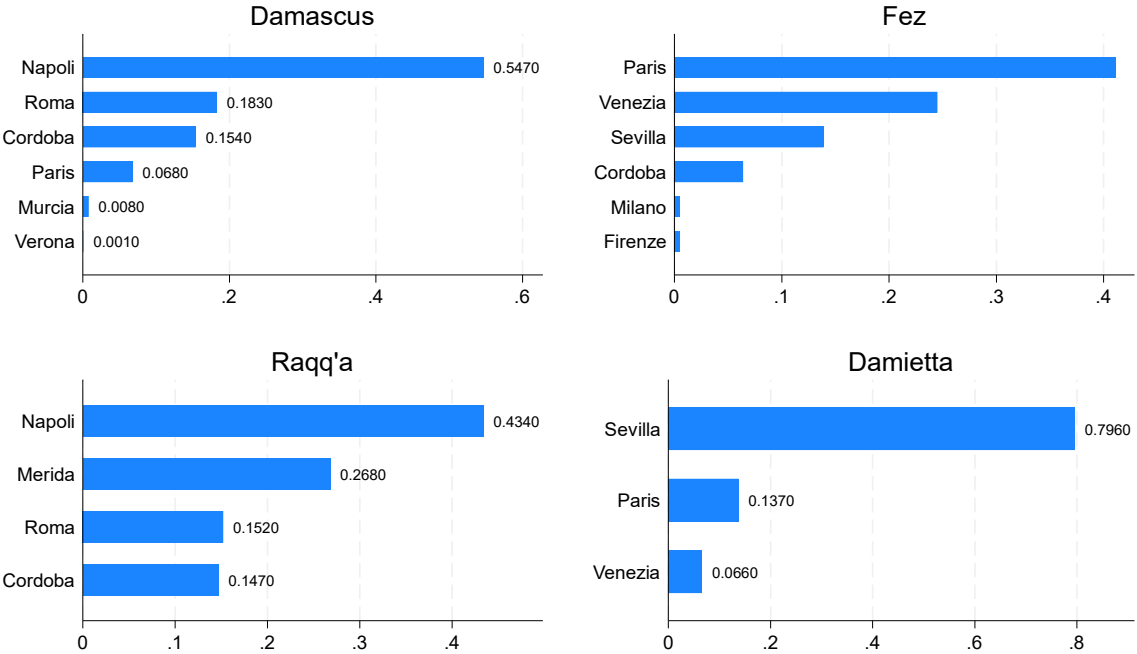

Figure 15: The composition of synthetic control groups for large cities

*5.3.2.2 Small city-level evidence*

Table 10 presents the evidence of the impact of the Shariatic turn on the population size for a subset of smaller cities. We have randomly selected four cities from our treatment sample, namely, Qus, Medina Al Fayyum, San'a and Mecca. For each city, we estimate the missing counterfactual scenario associated with the hypothetical absence of the Shariatic turn and the realization of western legal reforms. The evidence from the synthetic control analysis of the post-siege decline arguably indicates large-scale reduction of city size after the consolidation of the Shariatic turn but somewhat smaller paled in comparison with the large cities. Our synthetic control estimated gap between the observed city size and its synthetic peer is between -11,891 inhabitants in San'a and -51,697 inhabitants in Qus. Across all four small cities, the estimated overall effect of the Shariatic turn is statistically significant at 1% and is also evaluated at the end-of-sample period through non-parametric permutation of the Shariatic turn to the entire donor pool of Western European cities, which were affected by the legal revolution of the 12th and 13th century. Pointwise, our small-city synthetic control estimates put forth that the negative post-Shariatic turn tends to grow considerably over time. For example, the observed population size of Qus in the year 800 was around 10,000 inhabitants, and at the time only fourteen cities in Western Europe had greater population size than Qus. Our synthetic control estimates highlight a rapid divergence in population size after the Shariatic turn in the 13$^{th}$ century, which appears to be permanent. By 1600, the observed population size of Qus was zero. Our estimates imply that in the hypothetical absence of the Shariatic turn and its consolidation as well as the realization of Western European legal innovations, the population size of Qus could reach around 56,000 inhabitants. We arrive at a similar magnitude for the city of Medina Al Fayyum whereas our estimates for San'a are somewhat more modest. In particular, by 1600, the difference between the Sana's observed city size and its synthetic counterfactual is around 15,000 inhabitants. This implies that in the eventual absence of the Shariatic turn and realization of western growth oriented legal innovations, Sana's population size could increase up to 41,000 inhabitants, which would be the observed equivalent of cities such as Valladolid or Leiden, respectively. For Mecca, our counterfactual estimate for the year 1600 is around 53,000 inhabitants which is the observed equivalent of Amsterdam. Nevertheless, it should be noted that the estimated magnitudes for smaller cities are particularly smaller than their counterparts for the large cities. The entire set of post-Shariatic turn synthetic counterfactuals is reported in Figure 16.

Table 10: Synthetic control estimated long-term effect of the Shariatic turn in Islamic law on long-term evolution of city size in 4 cities, 800-1800

|  | Small-City Treatment Samples | | | |
| --- | --- | --- | --- | --- |
|  | (1) | (2) | (3) | (4) |
|  | Qus | Medina Al Fayyum | San'a | Mecca |
| ATET | -51.697*** | -37.678*** | -11.891*** | -36.013*** |
|  | (.0001) | (.00001) | (.0001) | (.0001) |
| Treatment effect by year |  |  |  |  |
| 1300 | -21.246 | -8.571 | -7.209 | -18.958 |
| 1400 | -16.698 | -5.785 | -8.888 | -15.091 |
| 1500 | -28.881 | -19.975 | -1.997 | -8.240 |
| 1600 | -56.465 | -47.946 | -15.810 | -30.691 |
| 1700 | -85.276 | -57.471 | -14.937 | -61.345 |
| 1800 | -101.62 | -87.044 | -22.507 | -81.753 |

Notes: the table reports synthetic control estimated difference between the observed city size and the synthetic counterfactual of the European cities unaffected by the Shariatic turn in Islamic law for the entire post-Shariatic turn period after 1200. Standard errors are adjusted for arbitrary heteroskedasticity and within-city serially correlated stochastic disturbances using finite-sample adjustment of the empirical distribution function through two-way error component model. Cluster-robust standard errors are reported in the parentheses. Time-specific treatment effect of the Shariatic turn are obtained based on the standard treatment permutation proposed by Abadie et. al. (2010) by obtaining the distribution of non-parametric p-values and the associated effect size magnitudes. Asterisks denote statistically significant ATET coefficients at 10 (%), 5% (**), and 1% (***), respectively.

**Figure 16**: Long-term effect of the Shariatic turn in Islamic law on city-level population size trajectories of small cities across the Middle East and North Africa, 800-1800

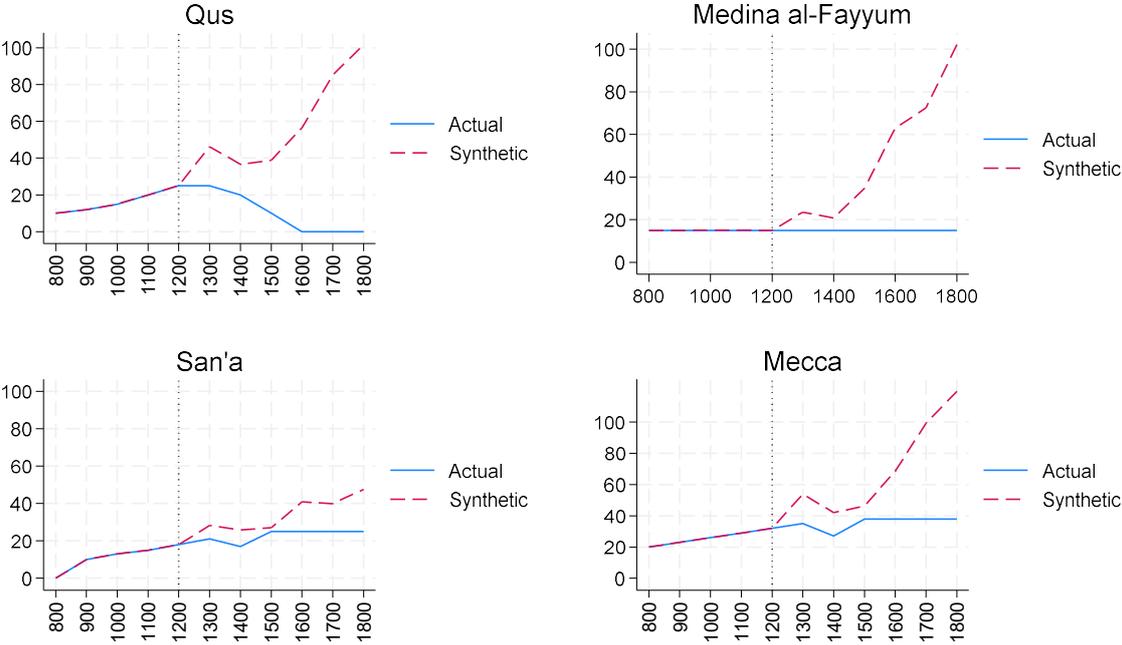

A detailed composition of synthetic control groups for the four considered cities is insightful and reflect the attributes of cities that plausibly reproduce the trajectories prior to 1300. For example, pre-1200 population size trajectory of Qus is best reproduced by a convex combination of the implicit characteristics of Paris, Cordoba, Cartagena, Murcia, Venice, Cologne, Verona and Naples. In varying proportions, these cities best reproduce the Qus's trajectory of population size prior to the siege of Baghdad. In addition, the population size trajectory of Medina Al Fayyum prior to the siege is best synthesized by a convex combination of the size attributes of Naples, Porto, Murcia, Rome, as well as Venice and Cordoba. Furthermore, Sana's synthetic control group consists of Venice, Porto and Salerno in addition to few others with minor weight share (i.e. Armagh, Badajoz, Huesca etc.) whereas Mecca's synthetic control group loads strongly on Cordoba (17%), Paris, and Cartagena, to a lesser extent, on Rome, Naples, Murcia and several others.

*5.3.2.3 Full-sample evidence*

Table 11 presents full treatment sample-level evidence and reports the estimated synthetic counterfactual city size trajectories in the eventual absence of the Shariatic turn and the realization of Western European legal reforms. The evidence is presented for four major groups of cities. The evidence invariably suggests large-scale city-level population losses in the post-Shariatic turn period. For instance, the average treatment effect of the Shariatic turn is around 10,000 fewer inhabitants for the cities in the Arab peninsula. The estimated ATET parameter is both large and statistically significant at 1%. In addition, cities in the Levantine and North African region appear to be most severely affected by the Shariatic turn. On average, synthetic counterfactuals indicate between 17,000 and 19,000 fewer inhabitants than their comparable Latin European control group. It should be noted that substantial effect heterogeneity is perceptible. Decomposing the effect by year unveils important disparities in the direction and size of the effect. For instance, the path of the diversion in pre-industrial city size in comparison to Western Europe appears to increase over time. Compared to the plausible Latin European control groups, the effect of the Shariatic turn in the cities on the Arab peninsula becomes statistically significant by the end of 16$^{th}$ century. By contrast, the effect of the Shariatic turn in North African cities becomes statistically significant at the conventional bounds immediately where the estimated gap for the Levantine cities becomes statistically significant by the end of the 15$^{th}$ century. The cities in Ottoman territories appear to be somewhat temporarily affected by the Shariatic turn given that the estimated gap is negative and statistically significant in the year 1300, and gradually dissipates up to the end-of-sample period. The full set of city size gaps has been estimated using the in-space treatment permutation approach based on full random sampling algorithm with a large number of treatment permutations ranging from 1.68 billion to 3.34 billion, respectively.

Table 11: Synthetic control estimated long-term effect of the Shariatic turn on long-term evolution of city size in Islamic countries compared with western Europe, 1300-1800

|  | Small-City Treatment Samples | | | |
|---|---|---|---|---|
|  | (1) | (2) | (3) | (4) |
|  | Arab | Levantine | North Africa | Ottoman-Anatolian |
| ATET | -10.742*** | -19.602*** | -17.373*** | 3.616 |
|  | (2.492) | (4.962) | (6.237) | (3.128) |
|  | Treatment effect by year | | | |
| 1300 | -3.327 | 1.659 | -10.672*** | -2.517** |
| 1400 | -4.684 | -2.652 | -3.457* | 0.725 |
| 1500 | -2.551 | -12.842*** | -10.926*** | 4.095* |
| 1600 | -9.508 | -34.828*** | -24.700*** | 12.161** |
| 1700 | -17.819* | -49.383*** | -37.114*** | 9.723** |
| 1800 | -26.567** | -55.296*** | -52.053*** | 0.631 |

| Treatment permutation method | Full-random sampling | Full random sampling | Full-random sampling | Full random sampling |
|---|---|---|---|---|
| number of placebo simulations | 1.68 billion | 2.03 billion | 3.45 billion | 3.34 billion |

Notes: the table reports synthetic control estimated difference between the observed city size and the synthetic counterfactual of the European cities unaffected by the Shariatic turn in 1200 for the entire post-Shariatic turn period after 1300. Standard errors are adjusted for arbitrary heteroskedasticity and within-city serially correlated stochastic disturbances using finite-sample adjustment of the empirical distribution function through two-way error component model. Cluster-robust standard errors are reported in the parentheses. Time-specific treatment effect of the Shariatic turn in Islamic law are obtained based on the standard treatment permutation proposed by Abadie et. al. (2010) by obtaining the distribution of non-parametric p-values and the associated effect size magnitudes. Asterisks denote statistically significant ATET coefficients at 10 (%), 5% (**), and 1% (***), respectively.

## 5.4  Discussion

Our triple-differences and synthetic control estimates unveil several noteworthy comparisons that necessitate a plausible context of interpretation. It should be noted that our results highlight the Islam law schools (i.e. *madāris*) as a source of long-term difference in the level of pre-industrial prosperity in comparison with (Latin) Europe. We show that the foundation of madrasas prior to the Shariatic turn is generally associated with more vibrant and dynamic urban development whereas the relationship between the density of madrasas and city-level population size becomes negative and statistically significant after the triumph of the Shariatic turn, which contrasts directly with a large, positive and robust contribution of law schools in European universities after 1100 to the pre-industrial urban growth.

In this respect, it should be noted that our results capture the long-term effect of the turn towards Sharia-based interpretation of law that began in the 9$^{th}$ century and radicalized in the 13$^{th}$ century with the Shariatic turn, that is the comprehensive sacralization of Islamic law. They do not show an influence of the Quran or Islamic religion on the trajectory of pre-industrial development before the Shariatic turn. The absence of a discernible downward pre-1300 trend in pre-industrial city population growth renders the influence of the Quran and Sunna on long-term development very limited up to the 13$^{th}$ century. The Sharia influence on the law in Islamic countries had not evolved suddenly but involved in a gradual process beginning in the 10$^{th}$ century when the political fragmentation most probably fomented and facilitated the comprehensive sacralization of law by jurists and expanded the influence of both Hadith and Sunnah in the interpretation of law. During the 13$^{th}$ century, the Sharia gained a decisive and comprehensive importance for the law. Since time-varying technology shocks common to all cities and the variation in the observable determinants of city size remains stable over time, negative and statistically significant post-1200 coefficients are indicative. The increasing density of madrasas, which reflect the influence of the new legal doctrine, negatively affects city size after 1200, whereas such effect is completely absent in the pre-Shariatic turn period. If anything, we find evidence of the small and positive contribution of medieval pre-Shariatic turn law schools on the respective size of cities. Although the lack of data that could be used to detect the prevalent schools of Islamic jurisprudence presents a limitation per se, existing evidence from the scholarly discussion invariably suggests that the more conservative school of thought (i.e. Hanafi) dominated the Islamic jurisprudence beginning in the $13^{th}$ century but not before.

Our findings thus indicate post-Shariatic turn foundation of madrasas as one of the legal roots of the institutional and economic underdevelopment of the Middle East prior to the industrial revolution (Kuran 2012). The foundation of madrasas after the turning point inflicted by Shariatic turn posits a stark discontinuity from the trajectory of legal development in the early period of Islamic golden age. Early medieval Islamic legal system emerged from a major legal transformation from fundamental texts and hadiths to a more systematic canonization of law. This particular impetus fostered the predictability of law and reduced transaction costs in economic and commercial exchange. Towards the peak of the $10^{th}$ century, Islamic law buttressed a constant adaptation to new development which enabled both long-distance trade and ensured the safety of the trade routes that made the economic exchange over the vast empire possible. Therein, the major strengths of the early Islamic law may be derived from the relative easiness of the enforceability of contracts over vast distances. Until the collapse of the political structure of the Abbasid empire and its fiscal administration in the early $11^{th}$ century, low-cost enforceability of contracts and both the flexibility and adaption of law to the internal dynamics and external shocks coupled with a relatively homogenous language and religion provided an important impetus to trade and economic growth that can be observed until the early $12^{th}$ century.

In the first half of the $12^{th}$ century, a more rigorous systematization of law did not entail reduced flexibility as many popular claims would suggest. Instead, our estimates inextricably highlight the more gradual advent of the Sharia-based consolidation through the change in the relationship between the greater density of madrasas and pre-industrial growth. Given the hypothesis that greater density of Islamic law schools does not predict more dynamic and vibrant urban development cannot be rejected, it would be superfluous to claim that Islamic law has evolved into a stagnant and backward body of evidence. Instead, what our estimates show is that the triumph of the Shariatic turn precipitated a long delay in the adoption of the more dynamic forms and facets of economic institutions that boosted the economic growth trajectories of Western Europe before and especially after the end of the Black Death. After the Shariatic turn, Islamic law scholars never introduced insurance contracts and financial markets and was perhaps a pivotal juncture point, precipitating the jump-start of English and Dutch economy in the late 16th and early 17th century (North 1991).[18] By delaying and scholastically rejecting the concept of insurance contracts as gambling with fate, both large-distance trade and capital markets in Islamic societies never evolved into a dynamic engine that would enable the formation of large-scale enterprises and institutional innovation capable of triggering a bold jump-start of more vibrant economic growth.

Our estimates also highlight high opportunity costs of foregone adoption of *pacta sunt servanda* which serves as one of the most fundamental principles of law. In Latin Europe, *pacta sunt servanda* facilitated the enforceability of all contracts which also allowed the design and construction of new forms of contracts by contracting parties. After 1200, our estimates indicate a strong acceleration of pre-industrial urban development in Western Europe that can be readily

---

[18] More specifically, North (1991, p. 129-130) discusses the virtuous interplay between financial markets and institutional innovation in North-Western Europe: "*The evolution of capital markets was critically influenced by the policies of the state, because to the extent that the state was bound by commitments that it would not confiscate asset, or in any way use its coercive power to increase uncertainty in exchange, it made possible the evolution of financial institutions and the creation of more efficient capital markets. The shackling of arbitrary behaviour of rulers and the development of impersonal rules that successfully bound both the state and voluntary organizations were a key part of the institutional transformation. The development of an institutional process by which government debt could be circulated became a part of the regular capital market, and the process that enabled a government debt to be funded by regular sources of taxation was a major step in the evolution of capital markets… It was the Netherlands, and Amsterdam specifically, that these diverse innovations and institutions were put together to create the predecessor of the efficient modern set of markets that make possible the growth of exchange and commerce. An open immigration policy attracted businessmen; efficient methods of financing long-distance trade were developed, as were capital markets and discounting methods in financial houses that lowered the cost of underwriting this trade, the development of techniques of spreading risk and transforming uncertainty into actuarial, ascertainable risk, the create of large-scale markets that allowed for lowering the costs of information, and the development of negotiable government indebtedness all were a part of this story.*"

associated with the cultural legal revolution (Berman 1985). Most of the impetus to the general enforceability of contracts can be readily attributed to the rise of learned law where these diverse legal innovations were taught and discussed. In Islamic law, the absence of *pacta sunt servanda* implied that more sophisticated forms of economic organization that allow for more efficient sharing of risk simply could not develop in the lieu of the wide-standing and persistent institutional void that made more advanced and complex forms or organization impossible. And third, post-1100 expansion of urban development in Latin Europe was possible due to the emergence of the corporation and its legal personhood. The latter triggered rapid deployment of new technologies, mobilize savings and channel them into large-scale investments, reinforcing a virtuous cycle of organizational innovation and capabilities for a more efficient use of technology. In turn, the concept of corporation and its legal personhood reflects a legal system suitable for complex and more sophisticated forms of economic and business organization where law schools are required to train and educate lawyers and related professions. By contrast, the Middle East economies were contingent on the financing of trade and production through Islamic partnerships. These partnerships were typically limited in terms of size and duration. The Islamic partnership differed markedly from the concept of corporation. It had no legal standing or personhood on its own. For instance, if one of the partners deceased before the contract was completed, the assets had to be liquidated and distributed to the partner's descendants. However, the Quran stipulates that two thirds of any estate are reserved for the partner's relatives. De facto, the intertwining of inheritance requirement with the partnership implied that successful and long-lasting large businesses were rare. They could seldom survive across generations, keep partnerships small in size and short-lived.[19] Such inhibitive influence of the legal institutions implied that institutional

---

[19] Kuran (2004, p. 87) discusses the economic drawbacks of Islamic partnership using an analogy of the prominent Egyptian entrepreneur in late 16th century (i.e. 1580-1625), Ismail Abu Taqiyya. *"A leading merchant of his time, Abu Taqiyya made a fortune by importing coffee, whose use was just beginning to spread across the Middle East… Anticipating Starbucks by several centuries, he also promoted coffee consumption by building scores of coffeehouses…Sensing a potential for dramatic market expansion, he financed sugarcane production, established refineries, and sold sugar both domestically and in the broader Mediterranean market… First, Abu Taqiyya operated through myriads of small and independent partnerships involving geographically dispersed people. Each partnership was based on a separate contract designed for narrowly defined purpose… These partnerships were short-lived, or they pooled limited resources, or both. Second, Abu Taqiyya's conglomerate did not outlast him… After his death, some of his associates took over certain components of his conglomerate… Although many of his coffeehouses probably lived on under different owners and new financial arrangements, his decades-old web of connections disappeared with him, and no person or organization inherited his reputation…His commercial capital got dissipated… His heirs did not maintain the conglomerate… The number of claimants must have mattered. Abu Taqiyya's heirs included eleven*

modernization was delayed and did not gain ground until the beginning of 19th century, and also stifled the process of economic growth onto a static path incapable of forging a technological and organizational breakthrough (Kuran 2020). Delayed adoption of the institutional reforms is reflected in the negative post-1200 madrasa density coefficients, mirroring a rapid growth acceleration in Latin Europe, catalysed by the legal innovation that made the institutional reforms possible.

# 6   Conclusion

This empirical paper contributes to the literature relating economic development to institutions, that is, to legal norms, social norms and substitute institutions. We compare Islamic societies in the Middle East and North Africa with the economic development in Western (i.e. Latin) Europe for the period from 800 to 1600. In both regions, fundamental legal changes occurred in the late Middle Ages which had a deep and profound impact on subsequent economic development. Legal historians agree that in Latin Europe after ca. 1100 the focus of the law became individual autonomy and responsibility.  These departures from family, clan and group-oriented law occurred both in the canon law of ecclesiastical courts as well as in secular courts, which has been dubbed a "big bang" for a longer lasting economic development of Western Europe.

The centrepiece of these changes was the introduction of the freedom of contract, which made all contracts enforceable, not only the numerus clausus of contracts as in classical Roman law. A corollary was the invention of the insurance contract, and the contractual reformulation of the legal person for organizing a company as a joint-stock corporation. Furthermore, the introduction of the sanctity of testators will improved the role of women as part of the labor force and society. These developments were accompanied with a systematization and scientific method in the newly established universities (i.e. *Verwissenschaftlichung*). One can safely say that this description is a short-hand of professional consensus.

---

*surviving children and four surviving wives. A few of them tried to consolidate their shares of the estate… Within a decade family squabbles, illness and additional deaths took their toll…When a partner died, the resulting inheritance claims were usually limited to his own kin…This would compound the costs of premature termination, making merchants and investors try even harder to keep their partnerships small and short-lived."*

Such consensus does not exist among Islamic scholars on the fundamental legal changes that occurred in the period between 1000 and 1100. It is undisputed among oriental scholars that in the first 400 years of the Golden Age of Islam, the Sharia had only a limited influence on legal practice, that many different philosophical and legal views determined this practice, that court decisions were ambiguous, and legal certainty was low. It is also undisputed that starting around the year 1000, the Islamic law was canonized and systematized vis-á-vis Sharia, Quran and the Sunna collection of writings of close contemporaries of prophet Mohammed on his legal opinions and precedences.

Opposing views exist, however, on the effect of this particular canonization. An older view maintains that the canonization was accompanied with an originalistic and very narrow interpretation of the law, which led to inflexibility, stickiness and incapability of legal innovation, producing detrimental long-term economic effects. An opposing view holds that the canonization increased the security of law, and consequently reduced transaction costs, promulgating positive economic effects. A related hypothesis entails a less straightforward view. It maintains that before the year 1000, Islamic law was relatively flexible with too little legal certainty. After the codification, legal uncertainty was removed but the law became too inflexible, thus missing the interior optimum between flexibility and certainty.

Our paper tests these hypotheses for Latin Europe as well as the Middle East and North Africa, using city population size as a rough proxy for pre-industrial economic development. For both regions, we use the number of universities with law schools in Western Europe, and the number of law schools (i.e. *madrasas*) in Islamic cities because the density of law schools informs about the number of lawyers, and consequently, about the effectiveness with which the existing law was practiced. The results of our study are clear for both regions. In Latin Europe, the number of universities with law departments is highly positively correlated with the overall city size. Moreover, we show that cities which lie in the vicinity of the university tend to have significantly higher population growth than other cities. The opposite is true for the Islamic law schools. The number of law schools after the canonization and the Shariatic turn in the 13$^{th}$ century is negatively related to the city size. Furthermore, the cities in the geographic vicinity of the

*madrasa*s have even lower city size. Our original and hand-collected data show that since the 13th century, the effectiveness of law expressed by the number of law schools and lawyers had a negative impact on economic development in the Middle East and North Africa. Also, our analysis supports the hypothesis that the big bang in European legal development from 1100 to 1200 had a large and positive long-term economic effect. It rejects the hypothesis that the systematization and canonization of Islamic law starting around the year 1000 had a net positive impact on trajectories of Middle Eastern economies. It supports the thesis that the canonization and sacralisation of Islamic law was combined with a narrow and originalist interpretation method of the Sharia texts, which cannot be changed because of their sacredness. None of the fundamental legal innovations in Latin Europe during the Middle Ages found its way into the Islamic legal order. We show that this combination had profound and long-lasting negative economic consequences for the development of the Middle East. The estimated impacts do not appear to be driven by omitted variables, and survive a long battery of robustness and mis-specification tests.

We conclude with a series of counterfactual estimates. More specifically, we ask how the city population size across the Islamic cities would have evolved if the systematization and canonization of law had been accompanied by legal reforms which took place around the same time in Latin Europe, most importantly, the introduction of the freedom of contract. Using this approach for each Islamic city, for which we have population data, an artificial version of this city is constructed from the attributes of a pool of Latin European cities, and compared with the actual trajectories. We show that under these conditions, almost all cities in North Africa and the Middle East would have experienced substantially higher population size, and therefore, economic development by the early 17th century.